\documentclass{aa}  
\usepackage{graphicx}
\usepackage{txfonts}
\usepackage{natbib}
\usepackage{amsmath}
\usepackage{color}
\usepackage{longtable}
\usepackage{lscape}
\usepackage{enumerate}
\usepackage{url}
\usepackage{mathtools}
\begin{document} 

   \title{The onset of energetic particle irradiation in Class 0 protostars\thanks{{\it Herschel} is an ESA space observatory with science instruments provided by European-led Principal Investigator consortia and with important participation from NASA.}}

   \subtitle{}

   \author{C. Favre
          \inst{1,2}
           \and
            A. L\'opez--Sepulcre\inst{3}
          \and
          C. Ceccarelli\inst{1,2}
            \and
   	C. Dominik \inst{4}
          \and          
	P. Caselli\inst{5}
          \and
	E. Caux\inst{6,7}
          \and       
          A. Fuente\inst{8}
          \and            
          M. Kama\inst{9}           
          \and          
	J. Le~Bourlot\inst{10}
          \and
   	B. Lefloch\inst{1,2}
          \and                 
	D. Lis\inst{10,11}
          \and
        T. Montmerle\inst{12}	
          \and
        M. Padovani\inst{13}	
          \and
     	C. Vastel\inst{6,7}
	          }

       \institute{Univ. Grenoble Alpes, IPAG, F--38000 Grenoble, France 
          \and      
          CNRS, IPAG, F--38000 Grenoble, France\\
              \email{cecile.favre@univ-grenoble-alpes.fr}
         \and         
	Institut de Radioastronomie Millim\'etrique, 300 rue de la Piscine, F--38406 Saint Martin d'H\`eres, France.
	\and
         Astronomical Institute "Anton Pannekoek", University of Amsterdam, Kruislaan 403, NL-1098SJ Amsterdam, Netherlands
         \and
  	Max-Planck-Institute for Extraterrestrial Physics (MPE), Giessenbachstr. 1, 85748 Garching, Germany
         \and
	Universit\'e de Toulouse, UPS-OMP, IRAP, Toulouse, France
         \and	
	CNRS, IRAP, 9 Av. colonel Roche, BP 44346, 31028, Toulouse Cedex 4, France     
         \and
	Observatorio Astron\'omico Nacional (OAN, IGN), Apdo 112, 28803 Alcal\'a de Henares, Spain 
          \and
         Leiden Observatory, P.O. Box 9513, NL--2300 RA, Leiden, The Netherlands
         \and      
         LERMA, Observatoire de Paris, PSL Research University, CNRS, Sorbonne Universit\'es, UPMC Univ. Paris 06, F--75014, Paris, France 
         \and
	Cahill Center for Astronomy and Astrophysics 301-17, California Institute of Technology, Pasadena, CA 91125, USA            
         \and
	Institut d’Astrophysique de Paris, 98bis Bd Arago, 75014 Paris, France
         \and
	INAF-Osservatorio Astrofisico di Arcetri, Largo E. Fermi, 5 -- 50125 Firenze, Italy
             }

   \date{Received September 15, 1996; accepted March 16, 1997}

 
%
%
  \abstract
   {The early stages of low-mass star formation are likely to be subject to intense ionization by protostellar energetic MeV particles. As a result, the surrounding gas is enriched in molecular ions, such as HCO$^{+}$ and N$_{2}$H$^{+}$. Nonetheless, this phenomenon remains poorly understood for Class 0 objects. Recently, based on Herschel observations taken as part of the key program Chemical HErschel Surveys of Star forming regions (CHESS), a very low HCO$^{+}$/N$_{2}$H$^{+}$ abundance ratio of about 3-4, has been reported toward the protocluster OMC-2~FIR4. This finding suggests a cosmic-ray ionization rate in excess of 10$^{-14}$ s$^{-1}$, much higher than the canonical value of $\zeta$ = 3$\times$10$^{-17}$ s$^{-1}$ (value expected in quiescent dense clouds).}
   {To assess the specificity of OMC-2~FIR4, we have extended this study to a sample of sources in low- and intermediate mass. More specifically, we seek to measure the HCO$^{+}$/N$_2$H$^{+}$  abundance ratio from high energy lines (J $\ge$ 6) toward this source sample in order to infer the flux of energetic particles in the warm and dense gas surrounding the protostars.}
   {We use observations performed with the Heterodyne Instrument for the Far--Infrared spectrometer on board the Herschel Space Observatory toward a sample of 9 protostars.}
   {We report HCO$^{+}$/N$_2$H$^{+}$ abundance ratios in the range of 5 up to 73 toward our source sample. The large error bars do not allow us to conclude whether OMC-2~FIR4 is a peculiar source. Nonetheless, an important result is that the measured HCO$^{+}$/N$_2$H$^{+}$ ratio does not vary with the source luminosity. 
At the present time, OMC-2~FIR4 remains the only source where a high flux of energetic particles is clearly evident. More sensitive and higher angular resolution observations are required to further investigate this process.}
   {}

   \keywords{stars: formation --
                stars: protostars --
                ISM: molecules --
                ISM: cosmic rays
               }

   \maketitle
%
%
\section{Introduction}
\label{sec:introduction}

\begin{sidewaystable*}
\caption{\label{tab1}Selected sources.}
\centering
\begin{tabular}{lllccccr}
\hline\hline
Source & R.A. (J2000) & Dec. (J2000) & D (pc) & L$\rm_{bol}$ (L$\rm_\odot$) & v$\rm_{LSR}$ (km~s$^{-1}$) &Ref & Herschel OBSIDs\\
\hline
\multicolumn{8}{c}{Low--mass}\\
\hline
VLA1623 & 16 26 26.4 &$-$24 24 30.0 & 120  & 1 & 3.8 &\small{1, 2, 3} & \small{1342249850, 1342250201, 1342250476, 1342250715, 1342250733,}\\
&&&&&&& \small{1342251210, 1342251211, 1342251248 and 1342251504} \\
L1527 &  04 39 53.9 & $+$26 03 10.0 & 140 & 2 & 5.9 & \small{4, 5, 6} & \small{1342249400, 1342249416, 1342249434, 1342249435, 1342249615,} \\
&&&&&&& \small{342249616, 342249861, 1342250193 and 1342250194 }\\
L1157--MM & 20 39 06.2 &  $+$68 02 22.0 &  325 & 11 & 2.6& \small{5, 6, 7, 8, 9} &\small{1342245336, 1342245337, 1342246461, 1342246462, 1342246499,}\\
&&&&&&& \small{1342245998, 1342245999, 1342246075 and 1342247020}\\
NGC1333--IRAS2 & 03 28 55.4  & $+$31 14 35.0  & 220 & 16 & 7.45 & \small{10, 11}& \small{1342248525, 1342248896, 1342248912, 1342248913, 1342249430,}\\
&&&&&&& \small{1342249431, 1342249604 and 1342249606}\\
\hline
\multicolumn{8}{c}{Intermediate--mass}\\
\hline
Serpens--FIRS~1 &  18 29 49.6 &$+$01 15 20.6  & 230 & 33 & 8 &\small{12, 13}&\small{1342243674, 1342251636, 1342251637, 1342253636, 1342253795,} \\
&&&&&&& \small{1342254436, 1342254437 and 1342268145}\\
L1641~S3~MMS~1 &  05 39 55.9 &  $-$07 30 28.0 & 500 & 67 & 5& \small{13}&\small{1342249619, 1342249863, 1342250183, 1342250184, 1342250679,} \\
&&&&&&& \small{1342251109, 1342251207, 1342251208 and 1342251507}\\
Cep~E--mm & 23 03 13.1 & $+$61 42 26.0 & 730 & 100 &$-$11&\small{ 6, 12, 13}&\small{1342246466, 1342246467, 1342246009, 1342246010, 1342246065,} \\
&&&&&&& \small{1342246074, 1342246333, 1342246337 and 1342247171}\\
IC1396N &  21 40 41.7 &  $+$58 16 12.8 &  750 & 150 &0.5 & \small{12, 13}&\small{1342245332, 1342245333, 1342245587, 1342246002, 1342246003,} \\
&&&&&&& \small{1342246464, 1342246465, 1342247022 and 1342247174}\\
NGC7129--FIRS2 & 21 43 01.7  & $+$66 03 23.6 & 1250 & 500 &$-$10 &\small{12, 13}&\small{1342243672, 1342245334, 1342245335, 1342246468, 1342246469,} \\
&&&&&&& \small{1342246000, 1342246001, 1342246001 and 1342247039}\\
\hline
\end{tabular}
\tablebib{ (1) \citet{Andre:1993}
(2) \citet{Loinard:2008}
(3) \citet{Maury:2012}
(4) \citet{Karska:2013}
(5) \citet{Kristensen:2012}
(6) \citet{Lopez-Sepulcre:2015}
(7) \citet{Straizys:1992} 
(8) \citet{Giannini:2001}
(9) \citet{Jorgensen:2007}
(10) \citet{Jorgensen:2004}
(11) \citet{Cernis:1990}
(12) \citet{Crimier:2010}
(13) \citet{Alonso-Albi:2010}.
}
\end{sidewaystable*}
%

\begin{table*}
\caption{\label{tab2}Spectroscopic line parameters.}
\centering
\begin{tabular}{lcrrr}
\hline\hline
Molecule	&	Transition	&	Frequency (MHz)	&	E$\rm_{up}$ (K)	&	S$\mu$$^{2}$ (D$^{2}$)	\\
\hline
HCO$^{+}$	&	6--5	&	535061.581	&	89.9	&	91.3	\\
HCO$^{+}$	&	8--7	&	713341.228	&	154.1	&	121.7	\\
HCO$^{+}$	&	11--10	&	980636.494	&	282.4	&	167.3	\\
HCO$^{+}$	&	12--11	&	1069693.891	&	333.8	&	182.5	\\
\hline
H$^{13}$CO$^{+}$	&	6--5	&	520459.884	&	87.4	&	91.3	\\
H$^{13}$CO$^{+}$ &	7--6	&	607174.646	&	116.6	&	106.5	\\
H$^{13}$CO$^{+}$	&	9--8	&	780562.812	&	187.3	&	136.9	\\
\hline
N$_{2}$H$^{+}$	&	6--5	&	558966.503	&	93.9	&	624.3	\\
N$_{2}$H$^{+}$	&	8--7	&	745209.868	&	160.9	&	832.3	\\
N$_{2}$H$^{+}$	&	11--10	&	1024443.025	&	295.1	&	1144.3	\\
\hline
\end{tabular}
\tablefoot{
We used the spectroscopic data parameters from \citet{Davies:1984}, \citet{Kawaguchi:1985}, \citet{Hirota:1988}, \citet{Botschwina:1993}, \citet{Lattanzi:2007} and \citet{Tinti:2007} for HCO$^{+}$, from \citet{Gregersen:2001}, \citet{Schmid-Burgk:2004} and \citet{Lattanzi:2007} for H$^{13}$CO$^{+}$ and, from \citet{Verhoeve:1990}, \citet{Havenith:1990}, \citet{Caselli:1995}, \citet{Amano:2005} and \citet{Pagani:2009} for N$_{2}$H$^{+}$. Note that all spectroscopic data are available from the Cologne Database for Molecular Spectroscopy molecular line catalog \citep[CDMS,][]{Muller:2005} at Splatalogue \citep[http://www.splatalogue.net,][]{Remijan:2007}.}
\end{table*}
%

Meteoritic materials conserve the traces of a violent past of the
early Solar System history. Specifically, the derived over-abundance of
short-lived radionuclides (SLRs) like $^{10}$Be provides evidence for a strong
irradiation ($\sim$ 10$^{19}$ -- 10$^{20}$ protons~cm$^{-2}$ in units of fluence) by energetic
($\ge$ 10 -- 20~MeV) particles \citep[e.g.][]{Gounelle:2013}. Various
theories have been evoked in the literature to explain the origin of
this energetic particle irradiation: a) galactic cosmic rays anchored
to the Solar prestellar clump magnetic field and focused in the Solar
Nebula \citep[e.g.][]{Desch:2004}, b) particles accelerated in the
atmosphere of the young Sun \citep{Bricker:2010}, c) particles
accelerated at the X-wind intersection \citep[e.g.][]{Gounelle:2006,Gounelle:2013}, 
and, finally, d) particles accelerated in a dense supersonic protostellar jet \citep{Padovani:2015,Padovani:2016}.

Unfortunately, it is difficult to distinguish between the above theories, since the irradiation period that happened in our own Solar System is over. If we could observe the process today in forming Sun--like
stars, then we might have more constraints to understand what happened
to our Solar System. To this end, we would need to discover
protostellar sources with signatures of energetic particle
irradiation.  Unfortunately, the direct detection of these energetic
particles is impossible. The problem is similar to that of detecting
the sources of acceleration of cosmic rays, as the latter are
scattered by the galactic magnetic fields and, consequently, have lost
the memory of their origin when they arrive on Earth. As in the cosmic
ray case, a way forward is to look for signatures of the interaction
of the energetic particles with the immediate surrounding material. It
turns out that two effects are, in principle, observable: i) an
enhanced $\gamma$-ray emission, caused by the interaction of $\geq 280$
MeV protons with the H-atoms \citep[e.g.][]{Hayakawa:1952,Stecker:1971} and,
ii) an enhanced ionization fraction, caused by the interaction of
0.1 -- 1~GeV particles \citep[e.g.][]{Indriolo:2010,Ceccarelli:2011}.

In practice, however, for sensitivity reasons, only the second effect
is observable in protostars. Also, since the source emitting the
energetic particles is, supposedly, embedded in the accreting
envelope, one needs to measure the ionization of the gas close to it,
namely in the inner warm and dense regions. One way to do it, is to
observe ions that have directly or indirectly been created by the energetic
particles, with relatively high upper
level energy transitions, in order to probe the dense and warm gas. Two ions satisfy
these two criteria: HCO$^+$ and N$_2$H$^+$. Indeed, both possess rotational transitions in the sub-millimeter with the appropriate
upper level energies ($\geq 100$ K), and they are created by the
reaction of H$_3^+$ (an ion that is almost directly created by the energetic particles or X-ray induced secondary electrons) with CO and N$_2$, respectively:
\begin{equation}\label{eq:1}
\rm{H}_3^+ ~~+~~ CO~~ \rightarrow ~~ HCO^+ ~~+~~ H_2
\end{equation}
\begin{equation}\label{eq:2}
\rm{H}_3^+ ~~+~~ N_2~~ \rightarrow ~~ N_2H^+ ~~+~~ H_2
\end{equation}

A first observation of the high lying transitions (with J$\geq 6$,
namely upper level energy $\geq$100 K) of HCO$^+$ and N$_2$H$^+$ was obtained  by \citet{Ceccarelli:2014} toward the protocluster OMC-2~FIR4 \citep{Shimajiri:2008,Lopez-Sepulcre:2013} by HSO ({\it Herschel Space Observatory}) within the Key Project CHESS
\citep{Ceccarelli:2010}. Unexpectedly, the
derived HCO$^+$/N$_2$H$^+$ abundance ratio turned out to be very
low, 3--4, when compared to that of gas in ``standard''
conditions  \citep[$\gg$ 10, e.g.][]{Sanhueza:2012,Hoq:2013}. When considering the formation
(Eqs. \eqref{eq:1} and \eqref{eq:2}) and destruction routes of both
molecules, it turns out that the only way to obtain such a low
HCO$^+$/N$_2$H$^+$ abundance ratio is when both molecules are
destroyed by electrons, which can only happen when the gas ionization
is dominated by \textit{i)} energetic particles \citep[which is the case of OMC-2~FIR4, see][]{Ceccarelli:2014} or \textit{ii)} X-ray irradiation \citep{Stauber:2005}, both of which interact with atomic and molecular hydrogen to produce H$_{3}^{+}$ and electrons.
Alternatively, a low HCO$^+$/N$_2$H$^+$ abundance ratio can be a result of the depletion of gaseous CO with respect to N$_2$ due to the CO freeze-out. Indeed, the CO gas-phase depletion by adsorption onto grains mantles would then decrease the HCO$^+$ abundance, and likely enhance that of N$_2$H$^+$ (see Section~\ref{sec:discussion}). However, this occurs only in cold ($\le$ 20--25~K) material, which is not the case in OMC-2~FIR4, since the temperature probed by the $J \geq 6$ transitions is $\ge$ 30 K.
Finally, a high abundance of water could also destroy HCO$^{+}$ and N$_2$H$^{+}$ \citep[e.g.][]{Stauber:2006} so that their ratio might decrease \citep[see also][]{Ceccarelli:2014}. Again, in the case of OMC-2~FIR4, the observed water abundance is too low for this explanation to be valid.
Therefore, the observed low ratio can only be explained by the presence of one or more embedded sources
emitting a large flux of energetic particles. Furthermore, the derived
dose of energetic particles is similar to that experienced by the 
young Solar System \citep[for further details, see][]{Ceccarelli:2014}.

These findings lead one to ask whether OMC-2~FIR4 is a
peculiar source or if other protostars do experience the same process. If the latter, what does the process depend on: age, luminosity, mass or
multiplicity? In order to address the above questions, we followed up with
a survey of the high-J HCO$^{+}$ and N$_2$H$^{+}$ HSO observations
toward a sample of low-- and intermediate--mass embedded protostars.
In Section~\ref{sec:observations}, we present the observations. Results are presented and discussed in Sections~\ref{sec:results}, \ref{sec:ratio} and \ref{sec:discussion}, with conclusions set out in Section~\ref{sec:conc}.

%
\section{Observations and data reduction}
\label{sec:observations}
%

\subsection{Source sample}\label{sec:source-sample}
Our survey is composed of a sample of nine well-known low- and
intermediate- mass embedded protostars that are listed in
Table~\ref{tab1} along with their respective coordinates, distances
from the Sun, luminosities and LSR velocities. The sources were
selected in order to cover a wide range of luminosities, from
1~L$\rm_\odot$ up to 500~L$\rm_\odot$ (see
Table~\ref{tab1}), and consequently, likely masses and evolutionary
stages.

\subsection{Observed frequencies}\label{sec:observed-frequencies}
As we aimed to derive accurate HCO$^{+}$/N$_{2}$H$^{+}$ abundance ratios
toward the inner dense and warm region of the envelope surrounding
the target sources, high transitions (6 $\le$ J $\le$ 12) of HCO$^{+}$
and N$_{2}$H$^{+}$ were selected. Furthermore, H$^{13}$CO$^{+}$
observations were performed to evaluate HCO$^{+}$ optical depth.  The
observed HCO$^{+}$, H$^{13}$CO$^{+}$ and N$_{2}$H$^{+}$ frequencies
are listed in Table~\ref{tab2} together with the spectroscopic line
parameters.  Observations of all the targeted lines (see
Table~\ref{tab2}) have been performed toward our source sample,
except for the HCO$^{+}$ (12--11) transition, which has
not been observed toward the following objects: NGC1333--IRAS2,
Serpens--FIRS~1 and NGC7129--FIRS2.

\subsection{Herschel HIFI observations}\label{sec:hersch-hifi-observ}
The observations were performed with the Heterodyne Instrument for the Far--Infrared (HIFI) spectrometer \citep{de-Graauw:2010} on board the Herschel Space Observatory \citep{Pilbratt:2010} between 2012 April and August as part of a Herschel Open Time. The data were obtained in fast chop dual beam switch (DBS program) mode pointed toward our sources (see coordinates in Table~\ref{tab1}) at a spectral resolution of 1.1~MHz. The HPBW lies in the range 18\arcsec up to 41\arcsec  (at 1.2~THz and 535~GHz, respectively). 

Data were exported to CLASS90 that is part of the GILDAS software \footnote{http://www.iram.fr/IRAMFR/GILDAS/} for reduction and analysis purposes. Calibration uncertainties are estimated to be less than 15$\%$ \citep{Roelfsema:2012}.
The continuum emission was fitted using a first--order polynomial and then subtracted from the scans. Then, the spectra that are reported in this study were -- for each source and each targeted line --  obtained  after "stitching" data from each scan. Also, noting that the line--widths of the observed lines are in the range of 1--4 km~s$^{-1}$ (see Section~\ref{sec:results}), the spectra were smoothed to a spectral resolution of 2.2~MHz. Finally, spectra that are shown in this paper are in units of the main beam temperature ($\rm T_\mathrm{MB}$). The intensity conversion from antenna temperature ($\rm T_\mathrm{A}^\ast$) to main beam temperature was done using the efficiencies (which include the frequency dependency) given by Michael Mueller and Willem Jellema (2014).\footnote{see \url{http://herschel.esac.esa.int/twiki/pub/Public/HifiCalibrationWeb/HifiBeamReleaseNote_Sep2014.pdf}.}

%
\section{Results}
\label{sec:results}

In the following (sub)sections, we present and describe in detail the results that were obtained with the HSO/HIFI data.

\subsection{Spectra}\label{sec:spectra}
\begin{figure}[h!]
\includegraphics[width=\hsize]{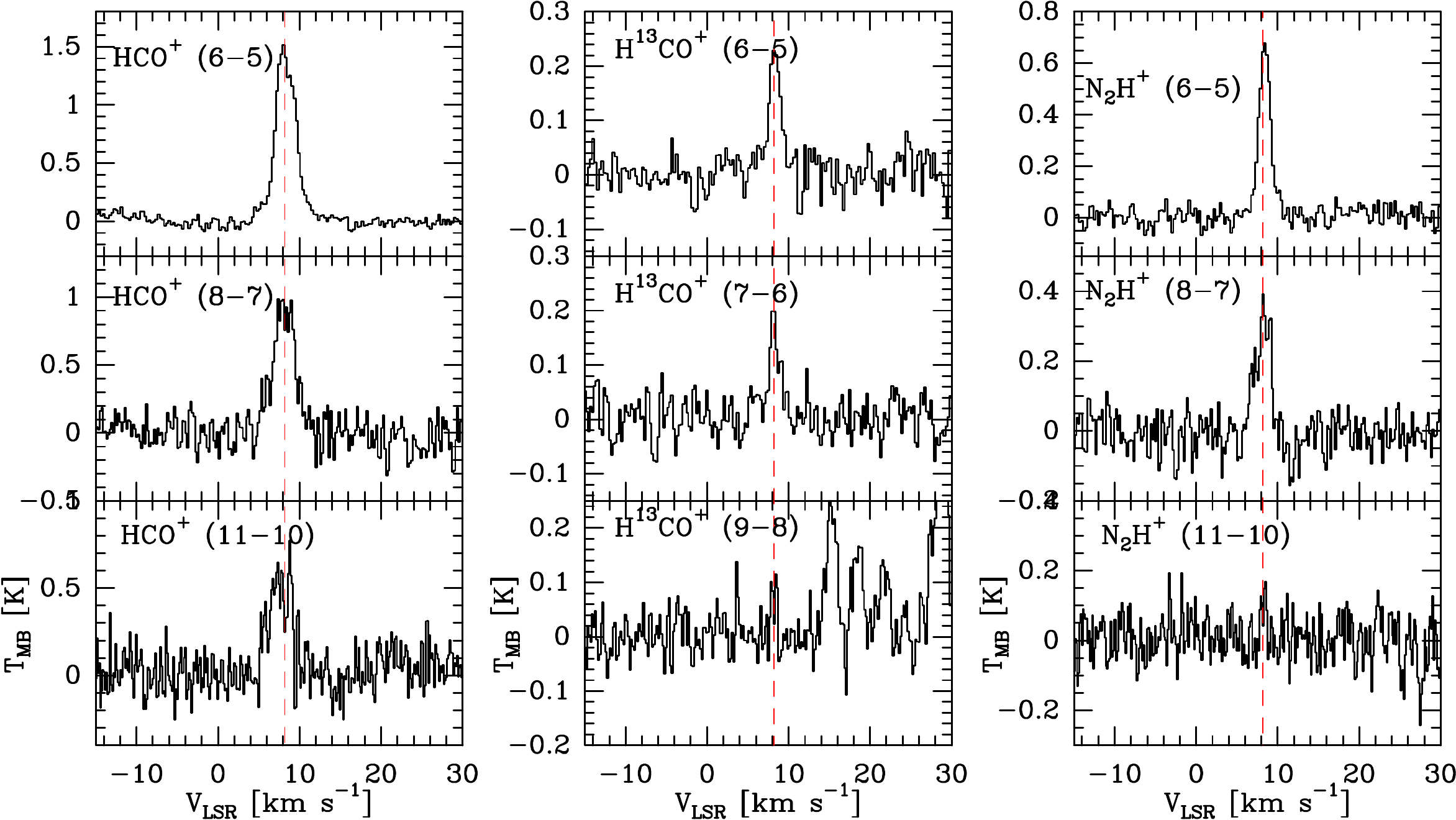}
\caption{Spectra observed toward Serpens--FIRS~1. Dashed red lines indicate a $v_{LSR}$ = 8.2~km~s$^{-1}$. The name of the observed transition is indicated on each plot. }
\label{fg1}
\end{figure}

\begin{figure}
\includegraphics[width=\hsize]{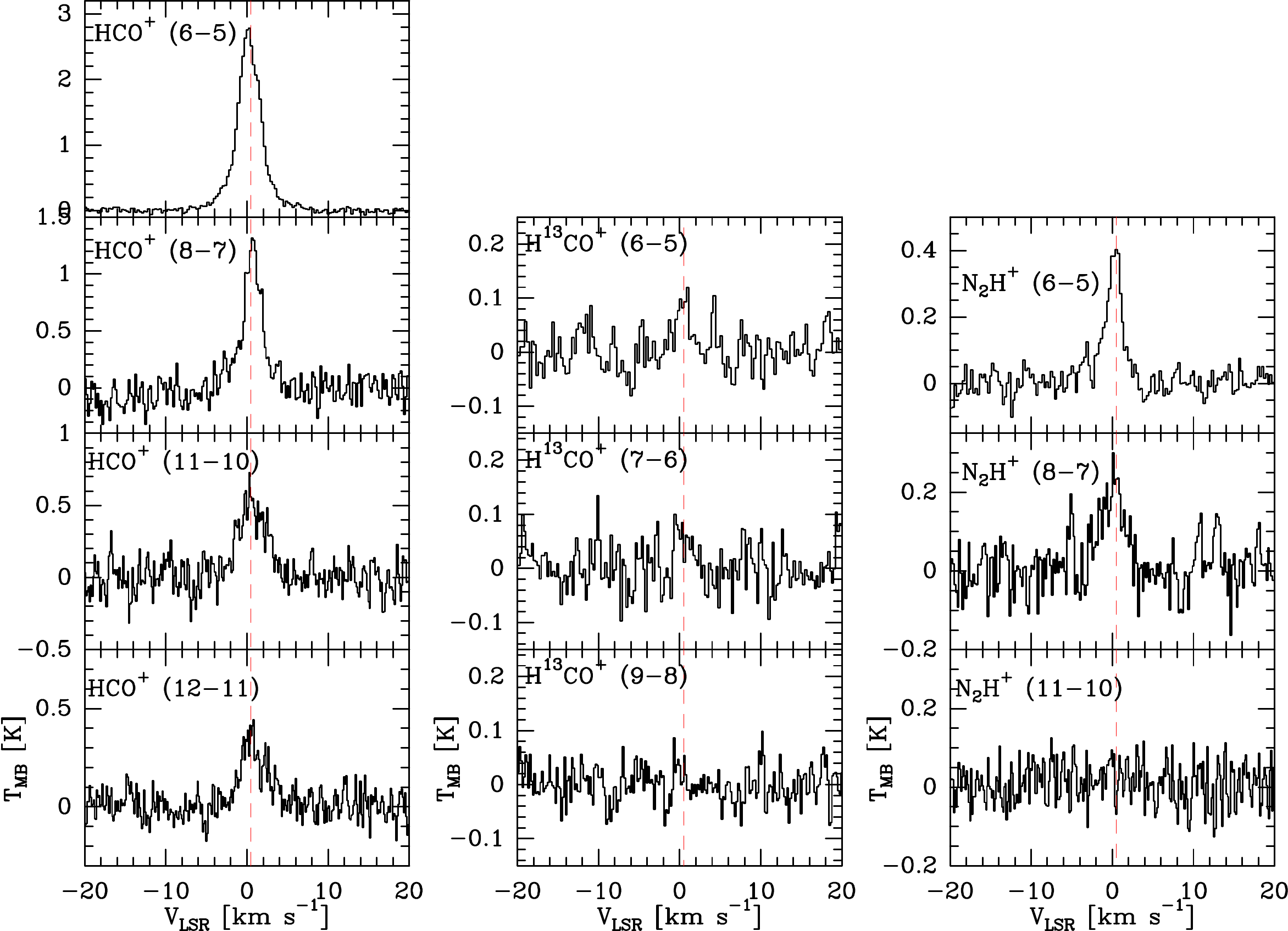}
\caption{Spectra observed toward IC1396N. Dashed red lines indicate a $v_{LSR}$ = 0.5~km~s$^{-1}$. The name of the observed transition is indicated on each plot. }
\label{fg2}
\end{figure}

Figures~\ref{fg1} and \ref{fg2} show the respective spectra of the HCO$^{+}$, H$^{13}$CO$^{+}$ and N$_{2}$H$^{+}$ transitions (see Table~\ref{tab2}) observed with Herschel toward the Serpens--FIRS~1 and IC1396N protostars and, Figures~\ref{fga1} to~\ref{fga7}, in Appendix~\ref{sec:app1}, display the spectra of the other observed sources.
For display purposes, the spectra have been smoothed to a spectral resolution of 4.4~MHz.
The bulk of the molecular emission appears to peak close to the systemic velocity of the targeted sources (see Table~\ref{tab1}).
The observed line parameters of the HCO$^{+}$, H$^{13}$CO$^{+}$ and N$_{2}$H$^{+}$ transitions toward our survey are summarized in Tables~\ref{tabB1} to~\ref{tabB9}, in Appendix~\ref{sec:app2}.
In the following analysis, we assume that the molecular emission line is \textit{i)} clearly detected if the peak intensity is greater than 3$\sigma$ and the integrated intensity, $\int$T$\rm_{MB}$dV, is greater than 5$\sigma$, \textit{ii)} "weakly" detected if the peak intensity is lower than the 3$\sigma$ level but  $\int$T$\rm_{MB}$dV $\ge$ 5$\sigma$ and, \textit{iii)} tentatively detected if both the peak intensity and $\int$T$\rm_{MB}$dV are at the 3$\sigma$ level. The measured integrated line intensities of HCO$^{+}$, H$^{13}$CO$^{+}$ and N$_{2}$H$^{+}$ in all the sources are listed in Table~\ref{tab3}. 

\begin{sidewaystable*}
\caption{\label{tab3}Integrated line intensities of HCO$^{+}$, H$^{13}$CO$^{+}$ and N$_{2}$H$^{+}$ toward our source sample.}   
\begin{tabular}{l|cccc|ccc|ccc}
\hline\hline
Source & \multicolumn{10}{c}{$\int$T$\rm_{MB}$dV (K~km~s$^{-1}$)} \\
 & \multicolumn{4}{c}{HCO$^{+}$} & \multicolumn{3}{c}{H$^{13}$CO$^{+}$}& \multicolumn{3}{c}{N$_{2}$H$^{+}$}\\ 
 & (6--5)	& (8--7) & (11--10) & (12--11) & (6--5) & (7--8) & (9--8)  & (6--5)	& (8--7) & (11--10) \\
\hline
VLA1623 &2.2(0.4)&0.7(0.2)&$\le$0.5& $\le$0.3&0.10(0.06)&$\le$0.2&$\le$0.3&0.16(0.06)&$\le$0.3&$\le$0.2  \\ 
L1527 & 1.26(0.09)&$\le$0.8 &$\le$1.0&$\le$0.6&$\le$0.3&$\le$0.4&$\le$0.3&$\le$0.3&$\le$0.5&$\le$0.4\\
L1157--MM & 0.63(0.09)&$\le$0.8&$\le$0.8&$\le$0.5&$\le$0.2&$\le$0.3&$\le$0.2&$\le$0.2&$\le$0.4&$\le$0.3 \\
NGC1333--IRAS2 &1.4(0.6)&1.5(0.3)&0.8(0.3)&--&0.09(0.06)&$\le$0.1&$\le$0.1&0.36(0.09)&$\le$0.6&$\le$0.6\\
Serpens--FIRS~1 & 3.8(0.6) & 3.2(0.3) & 1.9(0.3) & --& 0.41(0.09) & 0.29(0.09)& 0.08(0.06) & 1.21(0.09)& 0.7(0.1) & 0.10(0.08)\\
L1641~S3~MMS~1 & 0.9(0.4)&1.5(0.3)&1.6(0.3)&1.0(0.3)& $\le$0.2& $\le$0.1& $\le$0.1   &0.18(0.09)& $\le$0.3& $\le$0.2 \\
Cep~E--mm & 1.9(0.2)&1.3(0.4)&0.3(0.2)&$\le$0.9 &$\le$0.2&$\le$0.2&$\le$0.2 &0.11(0.06)&$\le$0.3&$\le$0.2 \\
IC1396N & 6.7(0.6) & 3.4(0.5)&2.24(0.05)&1.31(0.03)&0.21(0.10)&0.2(0.1) & $\le$0.4& 0.9(0.1)& 0.8(0.2)& $\le$0.6\\
NGC7129--FIRS2 &1.2(0.2)&1.6(0.4)&0.5(0.3) & --&$\le$0.5&$\le$0.5&$\le$0.4&0.34(0.09) &$\le$0.6&$\le$0.4\\
\hline
\end{tabular}
\tablefoot{
For further details, see Tables~\ref{tabB1} to~\ref{tabB9} in Appendix~\ref{sec:app2}. The 3$\sigma$ uncertainties that are given in brackets result from gaussian fits performed with the CLASS software. Finally, note that the HCO$^{+}$ (12--11) transition has not been observed toward NGC1333--IRAS2, Serpens--FIRS~1 and NGC7129--FIRS2 (see Sec.~\ref{sec:observed-frequencies}).}
\end{sidewaystable*}

\subsubsection{HCO$^{+}$}
All the targeted HCO$^{+}$ transitions (see Table~\ref{tab2}) are detected toward Serpens--FIRS~1, IC1396N, NGC1333--IRAS2, L1641~S3~MMS~1 and NGC7129--FIRS2. We note that the HCO$^{+}$ (6--5) transition is the only one detected toward L1527 and L1157--MM. Finally, we only report a detection of the HCO$^{+}$ (12--11) transitions in direction of the sources  L1641~S3~MMS~1 and IC1396N.
Incidentally, it is important to note that the line profile of HCO$^{+}$ (6--5) appears broadened (e.g., see, Fig.~\ref{fg2} toward IC1396N) by emission from different components within the Herschel beam. In this study, we only focus on the emission arising from the inner dense and warm region of the envelopes surrounding our source sample. We have therefore decomposed the observed profiles into two Gaussians: a broad one and a narrow one.  More specifically, we only analyse the emitting gas from the region/component in which we assume that  HCO$^{+}$ and N$_2$H$^{+}$ are both co-spatial, namely on the narrow component.  Therefore, the present study only reports in its Tables and Figures, the parameters associated with this region. 

\begin{table*}
\caption{\label{tab4}Beam dilution factor.}
\centering
\begin{tabular}{lccccc}
\hline\hline
Source & D (pc) & \multicolumn{4}{c}{Beam dilution factor, B} \\
 & & HCO$^{+}$(6--5)	&HCO$^{+}$ (8--7) & HCO$^{+}$(11--10) & HCO$^{+}$(12--11)  \\
\hline
VLA1623	&	120	&	0.82	&	0.89	&	0.94	&	0.95	\\
L1527	&	140	&	0.77	&	0.86	&	0.92	&	0.93	\\
L1157--MM	&	325	&	0.39	&	0.52	&	0.68	&	0.72	\\
NGC1333--IRAS2	&	220	&	0.58	&	0.71	&	0.82	&	0.85	\\
Serpens--FIRS~1	&	230	&	0.56	&	0.69	&	0.81	&	0.83	\\
L1641~S3~MMS~1	&	500	&	0.21	&	0.32	&	0.47	&	0.52	\\
Cep~E--mm	&	730	&	0.11	&	0.18	&	0.29	&	0.33	\\
IC1396N	&	750	&	0.11	&	0.17	&	0.28	&	0.32	\\
NGC7129--FIRS2	&	1250	&	0.04	&	0.07	&	0.12	&	0.15	\\
\hline
\end{tabular}
\end{table*}

 \subsubsection{H$^{13}$CO$^{+}$}
The H$^{13}$CO$^{+}$ (6--5) transition is detected toward the Serpens--FIRS~1, IC1396N, VLA1623 and NGC1333--IRAS2 sources. 
In addition, we report a detection and tentative detection of H$^{13}$CO$^{+}$ (7--6) in the direction of the Serpens--FIRS~1 and IC1396N intermediate--mass protostars, respectively. Incidentally, we tentatively detect H$^{13}$CO$^{+}$ (9--8) toward Serpens--FIRS~1.
Finally, we do not detect H$^{13}$CO$^{+}$ toward the following 5 sources: L1527, L1157--MM, L1641~S3~MMS~1, Cep~E--mm and NGC7129--FIRS2.

\subsubsection{N$_2$H$^{+}$} 
The N$_2$H$^{+}$ (6--5) transition is detected toward the following 6 protostars: Serpens--FIRS~1, IC1396N, VLA1623, NGC1333--IRAS2, L1641~S3~MMS~1 and NGC7129--FIRS2 and tentatively detected toward Cep~E--mm.
We also report the detection of N$_2$H$^{+}$ (8--7) in the direction of Serpens--FIRS~1 and IC1396N. The N$_2$H$^{+}$ (11--10) is tentatively detected toward Serpens--FIRS~1.
However, we do not detect N$_2$H$^{+}$ toward either L1527 and L1157--MM.\\

\noindent To summarize, the main observational results (see Table~\ref{tab3}), are as follows:
\begin{itemize}
\item The four HCO$^+$ transitions have been detected in the majority of the sources, with the lowest transition, $J=6-5$, detected in all of them.
\item On the contrary, the H$^{13}$CO$^+$ lines are mostly undetected, with even the $J=6-5$ line detected in only four out of the nine sources. 
\item Finally, the N$_{2}$H$^{+}$ $J=6-5$ line was detected in seven out the nine sources and the $J=8-7$ in only two of them.
\end{itemize}

\subsection{Spectral Line Energy Distribution (SLED) of HCO$^+$}
\label{sec:sled}

The sources observed in the present study lie at different distance from the Sun (see Table~\ref{tab1}). Beam dilution not only depends on the source size of the emitting region but in addition depends on the distance of the source. 
Under the assumption of an emitting region of 0.05~pc for each source, we can define the beam dilution factor, {\it B }, as follows \citep{Goldsmith:1999}:
\begin{equation}
 B = \frac{(0.05/d)^{2}}{(\theta_{b})^{2}+(0.05/d)^{2}},
\end{equation}
where d is the distance from the Sun (pc) and $\rm\theta_{b}$ the Herschel beam (\arcsec) , respectively. 
For each source, the respective derived {\it B} factor (see Table~\ref{tab4}) is used to correct for beam dilution the integrated line intensity of the HCO$^{+}$ lines.

\begin{figure}[h!]
\centering
\includegraphics[width=6cm]{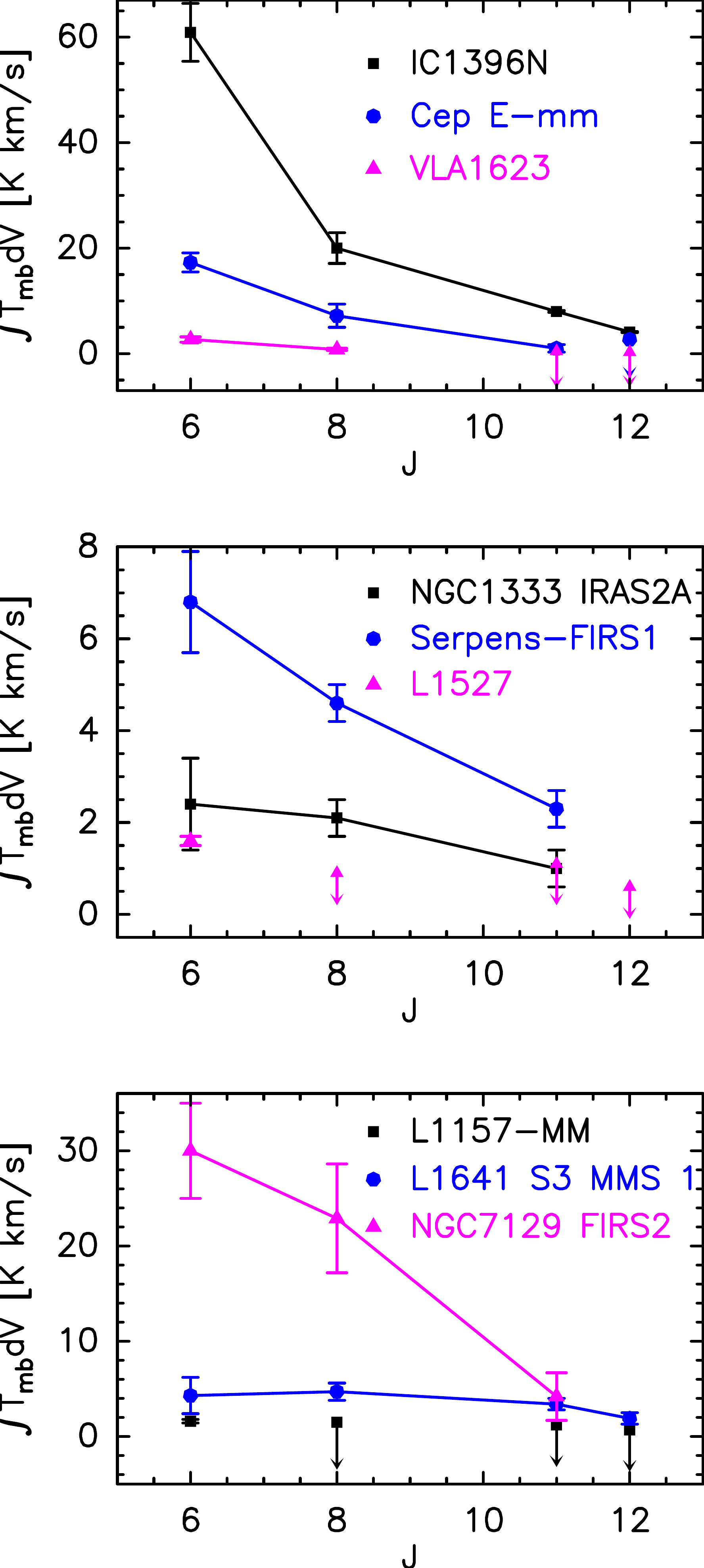}
\caption{SLED of HCO$^{+}$ as a function of the upper J transition toward the observed sources (see Table~\ref{tab1}). The name of the source is indicated in each panel.}
\label{fg3}
\end{figure}

Figure~\ref{fg3} shows the observed Spectral Line Energy Distribution (SLED) of HCO$^{+}$, corrected for beam dilution, as a function of the upper J transition toward all sources. Table~\ref{tab5} gives the measured integrated line intensity ratios, HCO$^{+}$ (6--5) / HCO$^{+}$ (8--7), toward our survey. 
It is immediately apparent that the HCO$^{+}$ $(6-5)/(8-7)$ line ratio suggests two classes of sources. Indeed, in VLA1623, IC1396N and perhaps Cep~E--mm, this ratio is about  3, whereas it is closer to unity (within the error bars) toward the other sources. This finding strongly suggests that the gas is warmer and denser in the latter group of sources. Nonetheless, a caveat of this approach is related to the opacities of the HCO$^{+}$ $(8-7)$ lines. Here, we assume that the emission is optically thin. However, if  $\rm \tau_{HCO^{+}(8-7)}$$\gg$1, the gas probed toward VLA1623 and IC1396N could be warmer.

Unfortunately, given the reduced number of detected lines in each source, we could not carry out a LVG analysis to constrain the density and temperature of the emitting gas, as was done in \citet{Ceccarelli:2014} for OMC-2~FIR4.

\begin{table}[h!]
\caption{\label{tab5}Measured integrated line intensity HCO$^{+}$ (6--5) / HCO$^{+}$ (8--7) ratios toward our source sample. }
\begin{tabular}{lc}
\hline\hline
Source & HCO$^{+}$ (6--5) / HCO$^{+}$ (8--7) \\
\hline
VLA1623 & 3.4(1.2) \\
L1527 & $>$1.8\\
L1157--MM & $>$ 1.1\\
NGC1333--IRAS2 & 1.1(0.5)  \\
Serpens--FIRS~1 & 1.5(0.3) \\
L1641~S3~MMS~1 & 0.9(0.4) \\
Cep~E--mm & 2.4(0.8) \\
IC1396N & 3.0(0.5) \\
NGC7129--FIRS2 &1.3(0.4) \\
\hline
\end{tabular}
\end{table}

%
\section{HCO$^{+}$/N$_2$H$^{+}$ abundance ratio determination}
\label{sec:ratio}

As explained in the Introduction, the goal of this article is to measure the HCO$^{+}$/N$_2$H$^{+}$ abundance ratio toward our sample of low-mass and intermediate--mass sources in order to infer the flux of energetic particles in the interior of the protostar envelopes. 
To derive this ratio, we used the integrated line intensity ratio of the HCO$^{+}$ (6-5) and N$_2$H$^{+}$ (6-5) lines, which are the main lines detected in the majority of the sources (see Table~\ref{tab3}).
This approach relies on the assumption that HCO$^{+}$ emits from the same region as N$_2$H$^{+}$. As these ($J=6-5$) transitions probe warm and dense gas and are chemically similar, this assumption is likely correct. More specifically, in the following sections, we obtain a robust estimate from the HCO$^{+}$ (6-5) over N$_2$H$^{+}$ (6-5) line intensity ratio toward each observed source. 
In addition, it is important to note that the measured integrated line intensity ratios do not suffer from beam dilution effect because the two targeted transitions have similar line frequencies and have been observed within the same beam.
Thus, since the two transitions have similar critical density and upper level energy (see Table~\ref{tab2}), the  intensity ratio conversion to an abundance ratio only assumes that the two species are co-existing in the telescope beam, once the ratio is corrected for the line opacity (see below). 
As a consequence, the obtained ratio represents an upper limit to the lowest HCO$^{+}$/N$_2$H$^{+}$ abundance ratio in the region.

\subsection{Line opacity}\label{sec:hco+-line-opacity}

As stated above, the HCO$^{+}$  abundance relative to N$_2$H$^{+}$ depends on the optical depth.
The observations of the H$^{13}$CO$^{+}$ lines, allow us to estimate the optical depth of the HCO$^{+}$ lines. 
To this end, we compare the integrated intensity of HCO$^{+}$ (6--5) to that of H$^{13}$CO$^{+}$ (6--5), when available (see Section~3.1.2 and Tables~\ref{tab3}, and~\ref{tabB1} to~\ref{tabB9}). 
From that, we estimate a flux ratio of 22, 16, 9 and 32 toward the VLA1623, NGC1333--IRAS2, Serpens--FIRS~1 and IC1396N protostars, respectively.
Thus, if we assume an isotopic ratio of $^{12}$C/$^{13}$C=68 for the local ISM \citep[see][]{Milam:2005}, the HCO$^{+}$ (6--5) line is optically thick in all these sources, with $\tau$ lying in the range $\sim$2 up to 8 (see Table~\ref{tab6}).

Regarding N$_2$H$^{+}$, it is important to state that our analysis hinges upon the assumption that the N$_2$H$^{+}$ (6--5) emission is optically thin. However, if $\rm \tau_{N_{2}H^{+}}$$\gg$1, the derived HCO$^{+}$/N$_2$H$^{+}$ abundance ratio should be an upper limit for the true HCO$^{+}$/N$_2$H abundance. In that instance, the true HCO$^{+}$/N$_2$H ratio would be smaller than the one reported in our study.

\begin{table}[h!]
\caption{\label{tab6}Measured integrated line intensity HCO$^{+}$ (6--5) / N$_2$H$^{+}$ (6--5) ratios toward our source sample. }
\begin{tabular}{lcc}
\hline\hline
Source & $\rm \tau_{HCO^{+}}$ & HCO$^{+}$ (6--5) / N$_2$H$^{+}$ (6--5) \\
\hline
VLA1623 &3.1& 42.5(30.1) \\
L1527 & -- & $>$4 \\
L1157--MM & --& $>$3 \\
NGC1333--IRAS2 & 4.5 & 17.0(12.1) \\
Serpens--FIRS~1 & 9.2 & 23.0(5.3)\\
L1641~S3~MMS~1 & -- & $>$5, $<$76 \\
Cep~E--mm  & -- & $>$17, $<$124\\
IC1396N & 2.9 & 15.9(7.8)\\
NGC7129--FIRS2 & -- &$>$4, $<$100\\
\hline
\end{tabular}
\end{table}

%
\subsection{HCO$^{+}$/N$_2$H$^{+}$ abundance ratio}\label{sec:hco+n_2h+-ratio}
\label{subsec:ratio}
\begin{figure}[h!]
\includegraphics[width=\hsize]{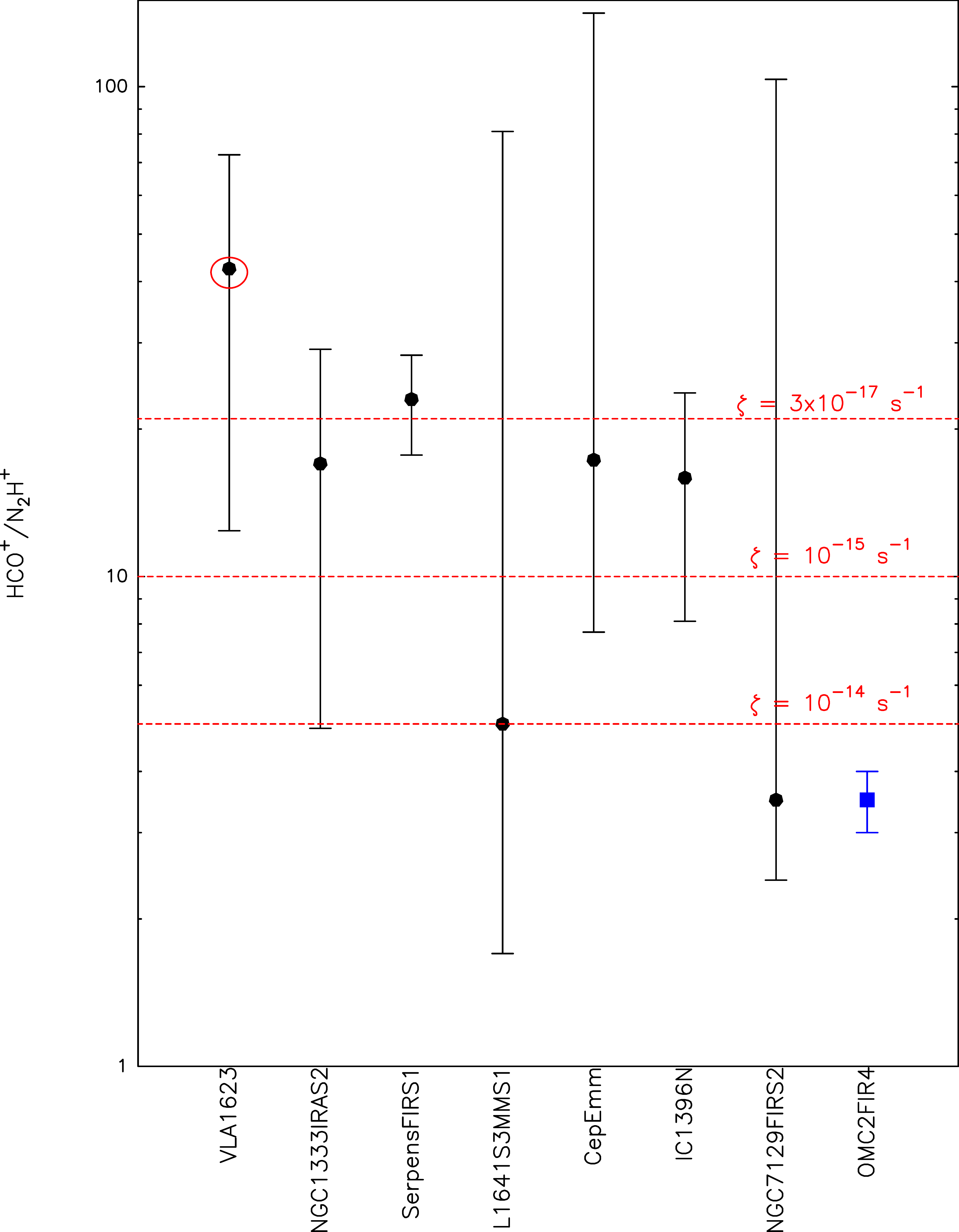}
\caption{Distribution of the HCO$^{+}$/N$_2$H$^{+}$ abundance ratios toward our source sample (see Table~\ref{tab1}). The sources are sorted by increasing luminosities. Full dots with error bars indicate the abundance ratios that have been corrected for the line opacity  when H$^{13}$CO$^{+}$ (6--5) observations are available (see Sections~\ref{sec:hco+-line-opacity} and \ref{sec:hco+n_2h+-ratio}). Red circle marks the colder source of our sample (based on the HCO$^{+}$ SLED, see Section~\ref{sec:sled}). The blue filled square shows the HCO$^{+}$/N$_2$H$^{+}$ ratio reported in direction of OMC-2~FIR4 by \citet{Ceccarelli:2014}. Finally, the red dashed lines show the predicted ratios that are expected for cosmic-rate ionization rates, $\zeta$, of 10$^{-14}$s$^{-1}$, 10$^{-15}$s$^{-1}$ and 3$\times$10$^{-17}$s$^{-1}$, for a gas temperature of 40~K and a n$\rm_{H_2}$  density of 2.5$\times$10$^{5}$~cm$^{-3}$ \citep[for further details see][]{Ceccarelli:2014}}.
\label{fg4}
\end{figure}

Figure~\ref{fg4} shows the distribution of the HCO$^{+}$/N$_2$H$^{+}$ (6--5) ratios, corrected for the HCO$^{+}$ line opacity when available (see Table~\ref{tab6}), that we derive toward our source sample. 
More specifically, the reported HCO$^{+}$/N$_2$H$^{+}$ abundance ratios for  VLA1623, NGC1333--IRAS2, Serpens--FIRS~1 and IC1396N are based on integrated line intensity ratios of the H$^{13}$CO$^{+}$ (6--5) over N$_2$H$^{+}$ (6--5), and assuming an isotopic $^{12}$C/$^{13}$C ratio of 68 \citep[see][]{Milam:2005}. Regarding L1641~S3~MMS~1, Cep~E--mm and NGC7129--FIRS2 protostars, the measured abundance ratio is only based on the ratio between the integrated line intensity of the HCO$^{+}$ (6--5) and N$_2$H$^{+}$ (6--5) transitions. In the case of $\tau \gg$1, we estimate  for these 3 sources, an upper limit of the HCO$^{+}$/N$_2$H$^{+}$ ratio (which is based on the upper limit of the $\int$T$\rm_{H^{13}CO^{+} (6-5)}$dV, see Table~\ref{tab3}) and,  include it in the error bars displayed in the Figure~\ref{fg4}.

It is noteworthy that the derived abundance ratios are given (in Fig.~\ref{fg4} and Table~\ref{tab6}) sorting the source by increasing luminosities and lie in the range of 4$^{+100}_{-1}$ up to 42$^{+31}_{-31}$.

%
\section{Discussion} \label{sec:discussion}

Numerous studies have investigated the HCO$^{+}$/N$_2$H$^{+}$ abundance ratio toward different astrophysical environments  \citep[e.g.][and references therein]{Turner:1977,Snyder:1977,Kim:2006,Lo:2009, Meier:2012,Sanhueza:2012,Ren:2014,Stephens:2015}. 
The latter is subject to variation according to the density of the region along with its chemical content (i.e. CO) and evolutionary stage. 

In particular, the HCO$^{+}$/N$_2$H$^{+}$ ratio is expected and measured to be low in early stages of star formation (T$<$10K). This results from the adsorption of CO onto grain mantles at low temperature. In that instance,  N$_2$H$^{+}$ is unlikely to be destroyed by CO, as suggested by Eq.~\ref{eq:3} \citep{Bergin:1997}:
\begin{equation}\label{eq:3}
\rm{N}_2H^+ ~~+~~ CO~~ \rightarrow ~~ HCO^+ ~~+~~ N_2
\end{equation}

On the contrary, in a more evolved stage, when CO is released in the gas-phase (typically T$>$20K), the HCO$^{+}$/N$_2$H$^{+}$ ratio is expected to be large, since the destruction of N$_2$H$^{+}$ by CO leads to an enhancement in HCO$^{+}$ \citep[see Eq.~\ref{eq:3} and][]{Bergin:1997}. 
Alternatively, the HCO$^{+}$/N$_2$H$^{+}$ ratio might be affected by the CO and N$_2$ abundances. On the one hand, the CO abundance may be lower that the canonical value of 10$^{-4}$: in protostellar envelopes \citep{Alonso-Albi:2010}, low--mass protostars \citep{Yildiz:2010,Anderl:2016} and, protoplanetary disk \citep[e.g.][]{Favre:2013}.
 On the other hand, N$_2$ may not be the main Nitrogen reservoir, Nitrogen being in the atomic-form and/or in ammonia ices \citep[e.g][and references therein]{Le-Gal:2014}.
 Finally, cosmic rays and X-rays might also impact this ratio \citep{Ceccarelli:2014,Bruderer:2009}, free electrons destroying both  HCO$^{+}$ and N$_2$H$^{+}$. Indeed, \citet{Ceccarelli:2014} have shown that CR--like energetic particles (cosmic-ray ionization rate $\zeta$ > 10$^{-14}$s$^{-1}$) can explain the very low and unusual HCO$^{+}$/N$_2$H$^{+}$ abundance ratio of about 3-4 that is observed toward the proto-cluster OMC-2~FIR4. Incidentally, \citet{Podio:2014} have also shown that a high value of the cosmic-ray ionization rate ($\zeta$ $\sim$ 3$\times$10$^{-16}$s$^{-1}$) reproduces the observed HCO$^{+}$ and N$_2$H$^{+}$ abundances toward a young protostellar outflow shock. 
As a matter of fact, X-rays may add up with cosmic rays in the production of H$_3^ +$ and free electrons, provided that the X-ray ionisation rate is high enough, which is not the case here, as  most of our source sample does not show significant X-ray emission (see Furusho et al. 2000 and the Chandra Data Archive).

\begin{figure}[h!]
\centering
\includegraphics[width=\hsize]{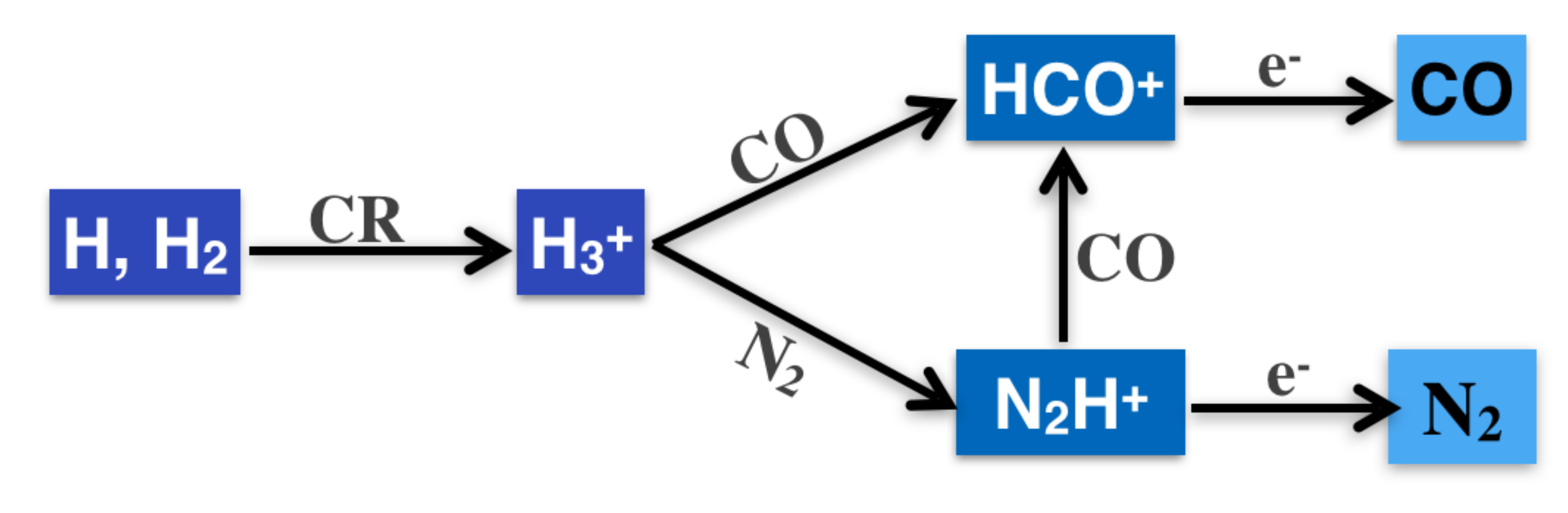}
\caption{Sketch of reactions that form and destroy HCO$^{+}$ and N$_2$H$^{+}$: when electrons do not dominate, CO is the main destroyer of N$_2$H$^{+}$. However, note that there are other branching ratios which give less abundant products that are not shown in this sketch. }
\label{fg5}
\end{figure}

Figure~\ref{fg5} illustrates and summarizes the main chemical HCO$^{+}$ and N$_2$H$^{+}$ destruction and formation pathways.

In that context, in Figure~\ref{fg4}, we compare the value derived toward OMC-2~FIR4 with the ones derived from our study.
In addition, we also report in Fig.~\ref{fg4} HCO$^{+}$/N$_2$H$^{+}$  ratios of 5, 10 and 21 that are predicted by the model of \citet{Ceccarelli:2014} for a gas temperature of 40~K and a n$\rm_{H_2}$ density of 2.5$\times$10$^{5}$~cm$^{-3}$, which are reasonable values for the gas probed by the Herschel observations presented in this work and, for a cosmic-ray ionization rate, $\zeta$, of 10$^{-14}$s$^{-1}$, 10$^{-15}$s$^{-1}$ and 3$\times$10$^{-17}$s$^{-1}$, respectively.
Higher HCO$^{+}$/N$_2$H$^{+}$ ratios (for example a value of 30) would correspond to lower $\zeta$ and conversely, lower HCO$^{+}$/N$_2$H$^{+}$ ratios would correspond to higher $\zeta$. We note that the case of L1641 S3 MMS 1 is interesting as the region where this source is lying contains X-ray sources (see Pillitteri et al. 2013) which may contribute to the ionization rate. Unfortunately, the error bars in our sample are relatively large and do not allow to draw firm conclusions on enhanced $\zeta$ of X-rays in any source.  On the contrary, VLA1623, Serpens~FIRS1, IC1396N and CepE do not show evident signs of large $\zeta$. This result comes as no surprise in the case of VLA1623 and IC1396N, since their HCO$^{+}$ SLEDs (see Figure~\ref{fg3}) are strongly suggesting cold sources. 
Nevertheless, one notable feature of Figure~\ref{fg4} is that the observed HCO$^{+}$/N$_2$H$^{+}$ ratio does not increase or decrease with increasing luminosity. 

We conclude that, at the present time, OMC-2~FIR4 is the only source where a high flux of energetic particles is clearly evident. Unfortunately, the statistics are small, so that we cannot infer what makes OMC-2~FIR4 special in that respect. More sensitive observations, especially toward NGC7129--FIRS2 and L1641~S3~MMS~1 (see pattern of the HCO$^{+}$ SLED along with the obtained HCO$^{+}$/N$_2$H$^{+}$  ratio in Figures~\ref{fg3}~and~\ref{fg4}), and in a much larger sample are necessary to say more.

%
\section{Conclusions}
\label{sec:conc}
We have investigated the HCO$^{+}$/N$_2$H$^{+}$ abundance ratio toward a sample of low-- and intermediate--mass protostars, through observations performed with the HIFI instrument on board the Herschel Space Observatory. Our study is based on the analysis of high J transitions (6 $\le$ J $\le$ 12) of the HCO$^+$, N$_{2}$H$^{+}$ and H$^{13}$CO$^{+}$ ions.

All the targeted HCO$^+$ transitions have been detected in the majority of the sources.  More specifically, the lowest lying transition (6-5) is detected toward all the 9 surveyed sources for HCO$^{+}$ and, in 7 of them for N$_{2}$H$^{+}$. Regarding H$^{13}$CO$^{+}$, the (6-5) line is only detected toward four out of the nine sources. Assuming a $^{12}$C/$^{13}$C isotopic ratio of 68, we report a HCO$^{+}$/N$_2$H$^{+}$ abundance ratio  in the range of 4$^{+100}_{-1}$ up to 42$^{+31}_{-31}$. The latter might even be lower depending on the opacity of the N$_2$H$^{+}$ (6--5) transition.
Incidentally, a salient result is that the measured HCO$^{+}$/N$_2$H$^{+}$ ratio does not increase or decrease with increasing luminosity, which suggests that UV radiation does not play a major role in the HCO$^{+}$/N$_2$H$^{+}$ abundance ratio.
However, our  measurements have large error bars that prevent us from determining whether OMC-2~FIR4 is the only source where a high flux of energetic particles is observed. Therefore, further sensitive and high angular resolution observations toward a much larger source sample are necessary \textit{i)} to ascertain whether OMC-2~FIR4 is a peculiar source and, \textit{ii)} to accurately infer the flux of energetic particles at the interior of the protostar envelopes if applicable.

%
%
\begin{acknowledgements}
Support for this work was provided by the French space agency CNES. MP acknowledges funding from the European Unions Horizon 2020 research and innovation programme under the Marie Sk\l{}odowska-Curie grant agreement No 664931. Support for this work was also provided by NASA (Herschel OT funding) through an award issued by JPL/Caltech. 
This paper makes use of Herschel/HIFI data. 
Herschel is an ESA space observatory with science instruments provided by European-led principal Investigator consortia and with important participation from NASA. HIFI has been designed and built by a consortium of institutes and university departments from across Europe, Canada, and the United States under the leadership of SRON Netherlands Institute for Space Research, Groningen, The Netherlands and with major contributions from Germany, France, and the U.S. Consortium members are: Canada: CSA, U. Waterloo; France: CESR, LAB, LERMA, IRAM; Germany: KOSMA, MPIfR, MPS; Ireland: NUI Maynooth; Italy: ASI, IFSI-INAF, Osservatorio Astrofisico di Arcetri-INAF; Netherlands: SRON, TUD; Poland: CAMK, CBK; Spain: Observatorio Astron\'omico Nacional (IGN), Centro de Astrobiolog\'ia (CSIC-INTA); Sweden: Chalmers University of Technology–MC2, RSS and GARD, Onsala Space Observatory, Swedish National Space Board, Stockholm Observatory; Switzerland: ETH Zurich, FHNW; USA: Caltech, JPL, NHSC. 
\end{acknowledgements}

%
%
\bibliographystyle{aa}

%
%
\Online
\begin{appendix}

%
%
\section{HCO$^{+}$, H$^{13}$CO$^{+}$ and N$_{2}$H$^{+}$ toward our sample of intermediate and low-mass protostars}
\label{sec:app1}

\begin{figure*}
\includegraphics[width=15cm]{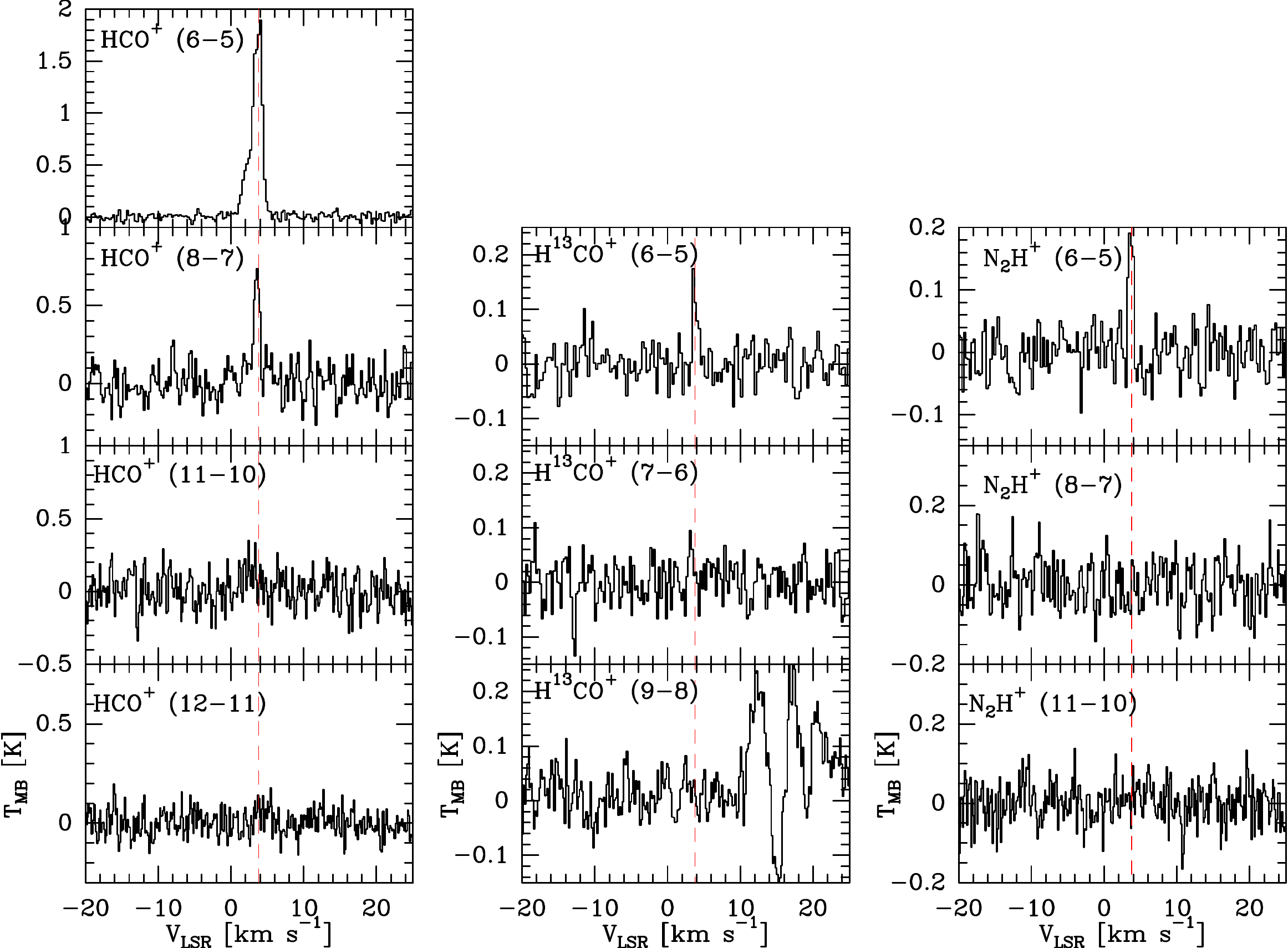}
\caption{Spectra observed toward VLA1623. Dashed red lines indicate a $v_{LSR}$ = 3.8~km~s$^{-1}$. The observed transition is indicated on each plot.}
\label{fga1}
\end{figure*}

\begin{figure*}
\includegraphics[width=15cm]{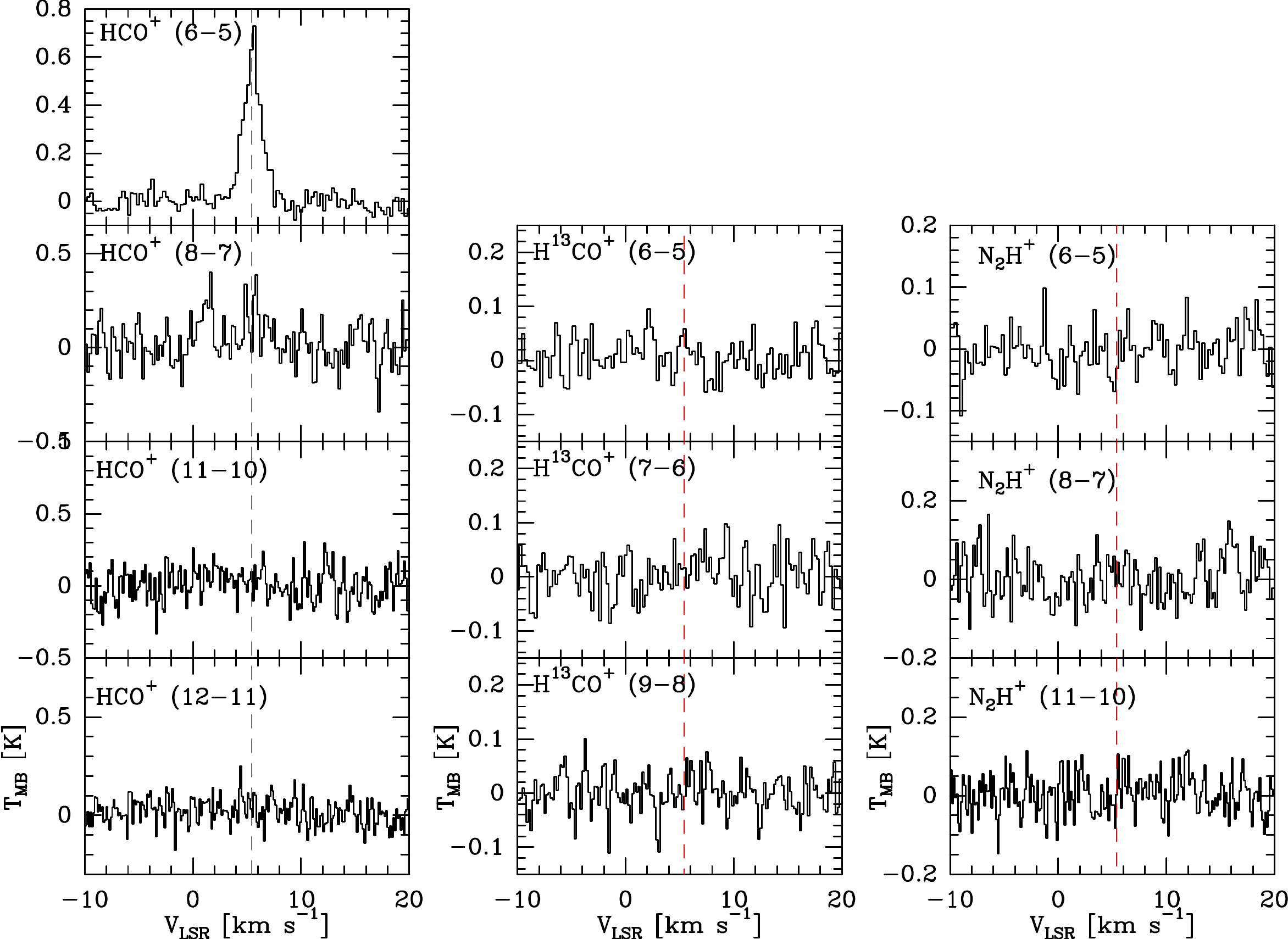}
\caption{Spectra observed toward L1527. Dashed red lines indicate a $v_{LSR}$ = 5.4~km~s$^{-1}$. The observed transition is indicated on each plot.}
\label{fga2}
\end{figure*}

\begin{figure*}
\includegraphics[width=15cm]{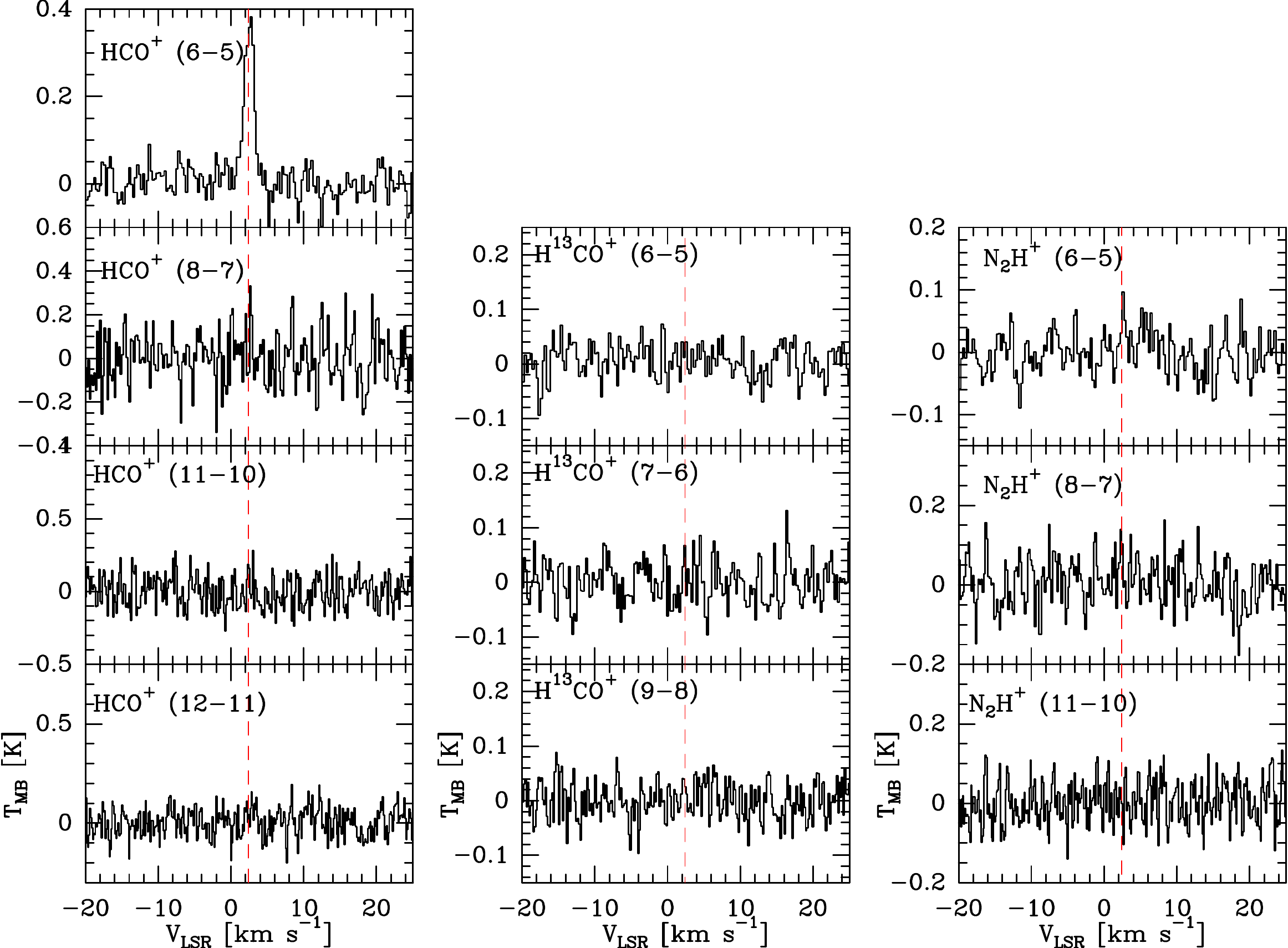}
\caption{Spectra observed toward L1157--MM. Dashed red lines indicate a $v_{LSR}$ = 2.4~km~s$^{-1}$. The observed transition is indicated on each plot.}
\label{fga3}
\end{figure*}

\begin{figure*}
\includegraphics[width=15cm]{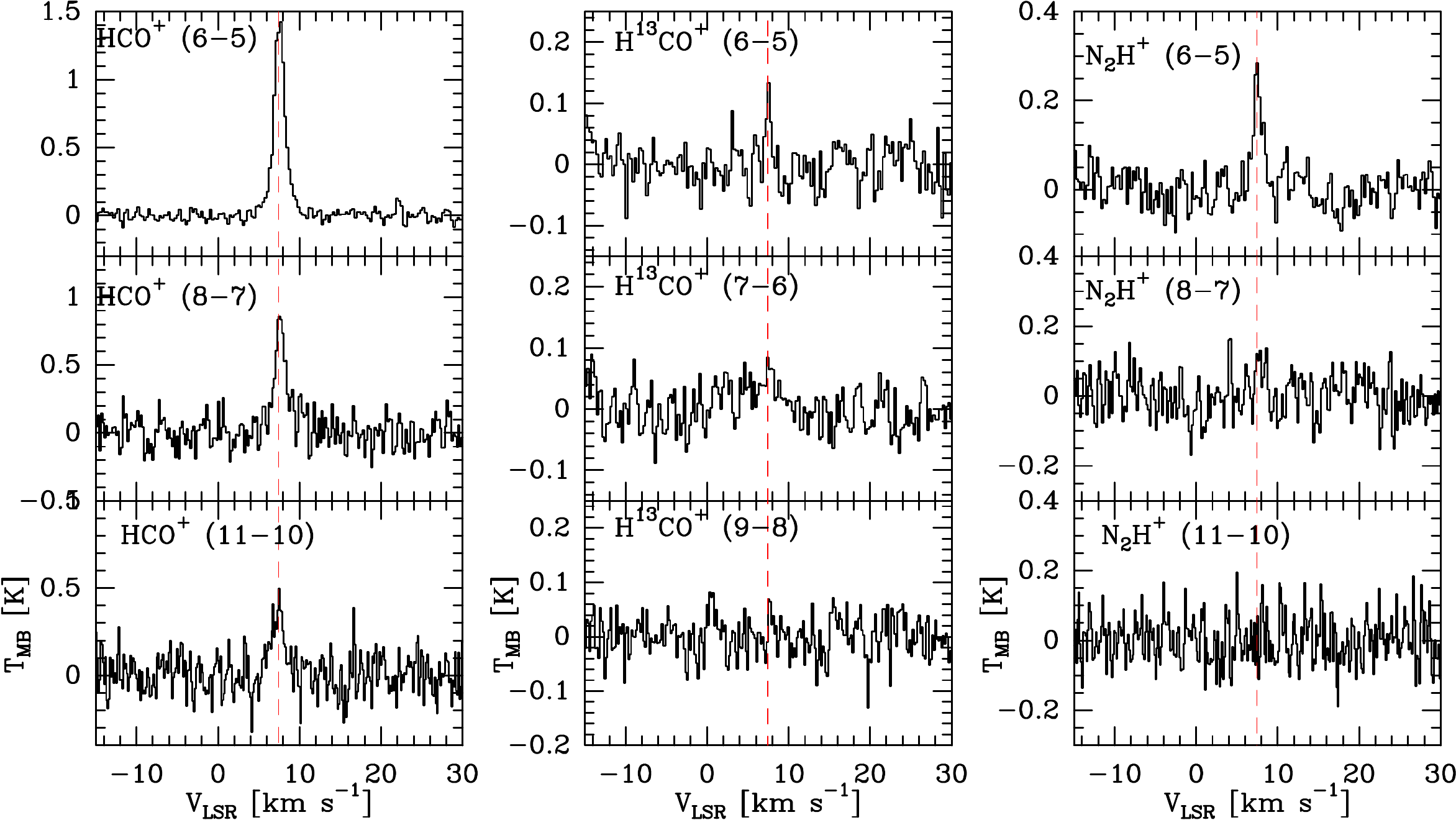}
\caption{Spectra observed toward NGC1333--IRAS2. Dashed red lines indicate a $v_{LSR}$ = 7.45~km~s$^{-1}$. The observed transition is indicated on each plot.}
\label{fga4}
\end{figure*}

\begin{figure*}
\includegraphics[width=15cm]{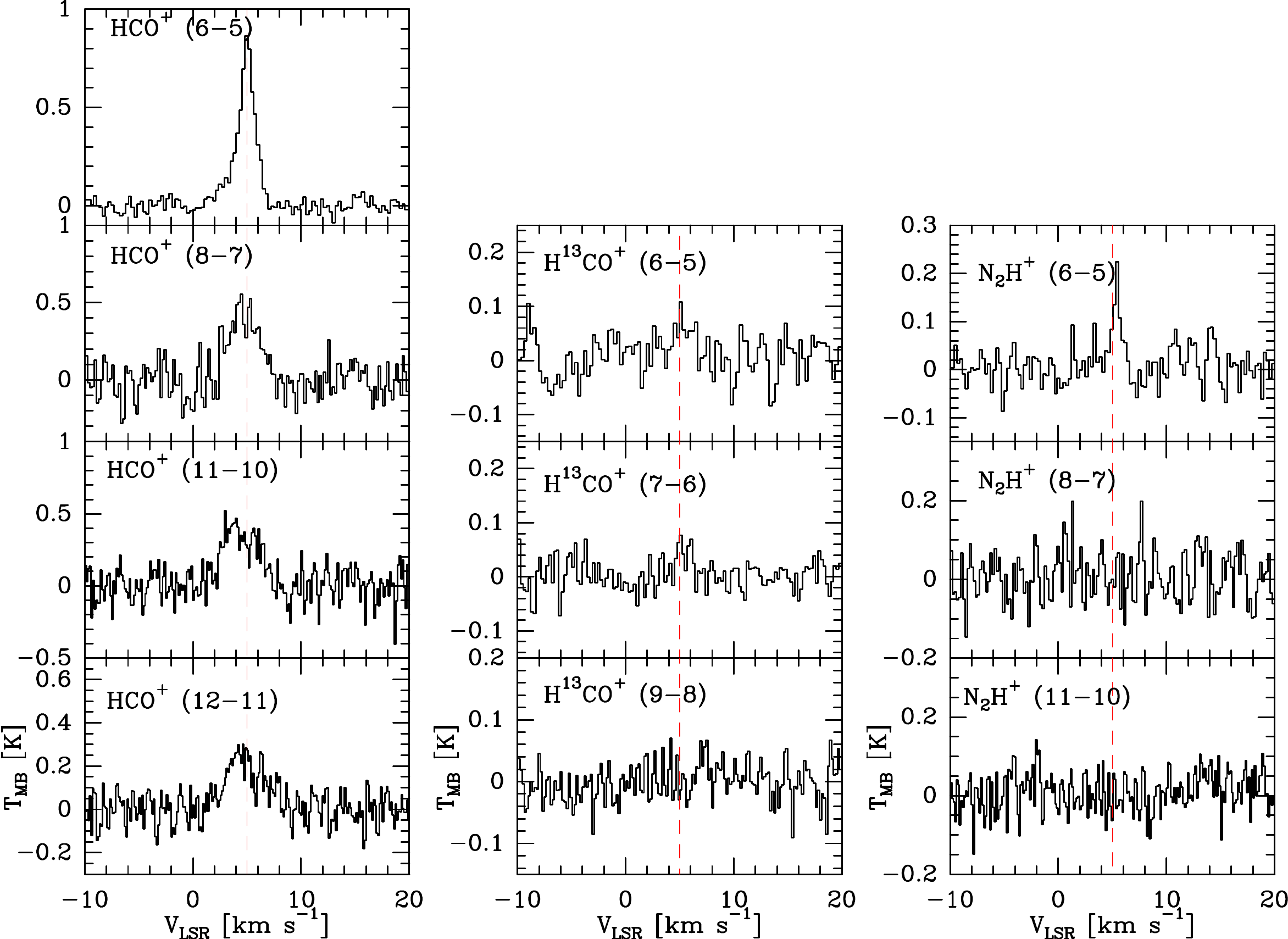}
\caption{Spectra observed toward L1641~S3~MMS~1. Dashed red lines indicate a $v_{LSR}$ = 5.0~km~s$^{-1}$. The observed transition is indicated on each plot.}
\label{fga5}
\end{figure*}

\begin{figure*}
\includegraphics[width=15cm]{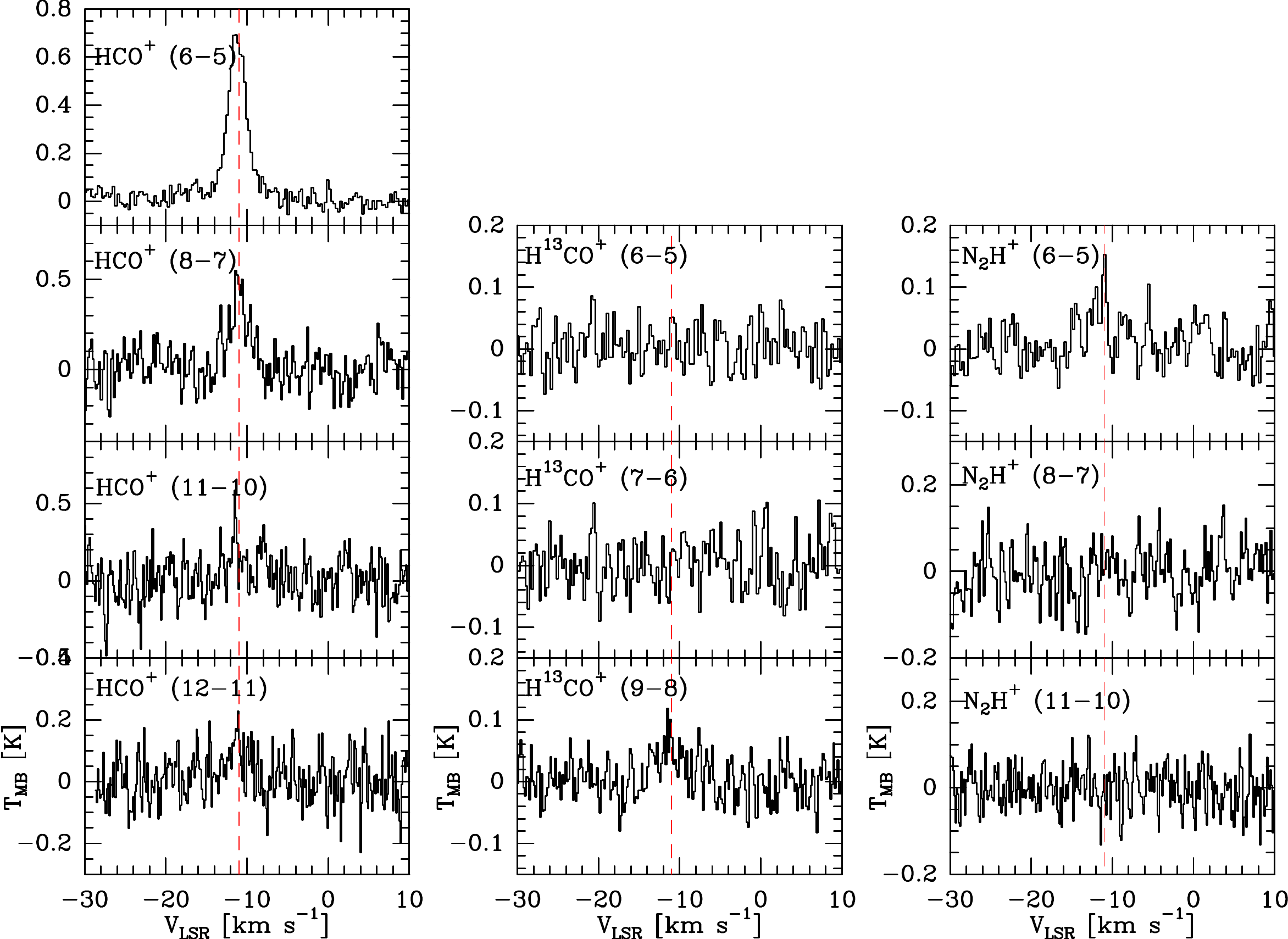}
\caption{Spectra observed toward Cep~E--mm. Dashed red lines indicate a $v_{LSR}$ = $-$11.0~km~s$^{-1}$. The observed transition is indicated on each plot.}
\label{fga6}
\end{figure*}

\begin{figure*}
\includegraphics[width=15cm]{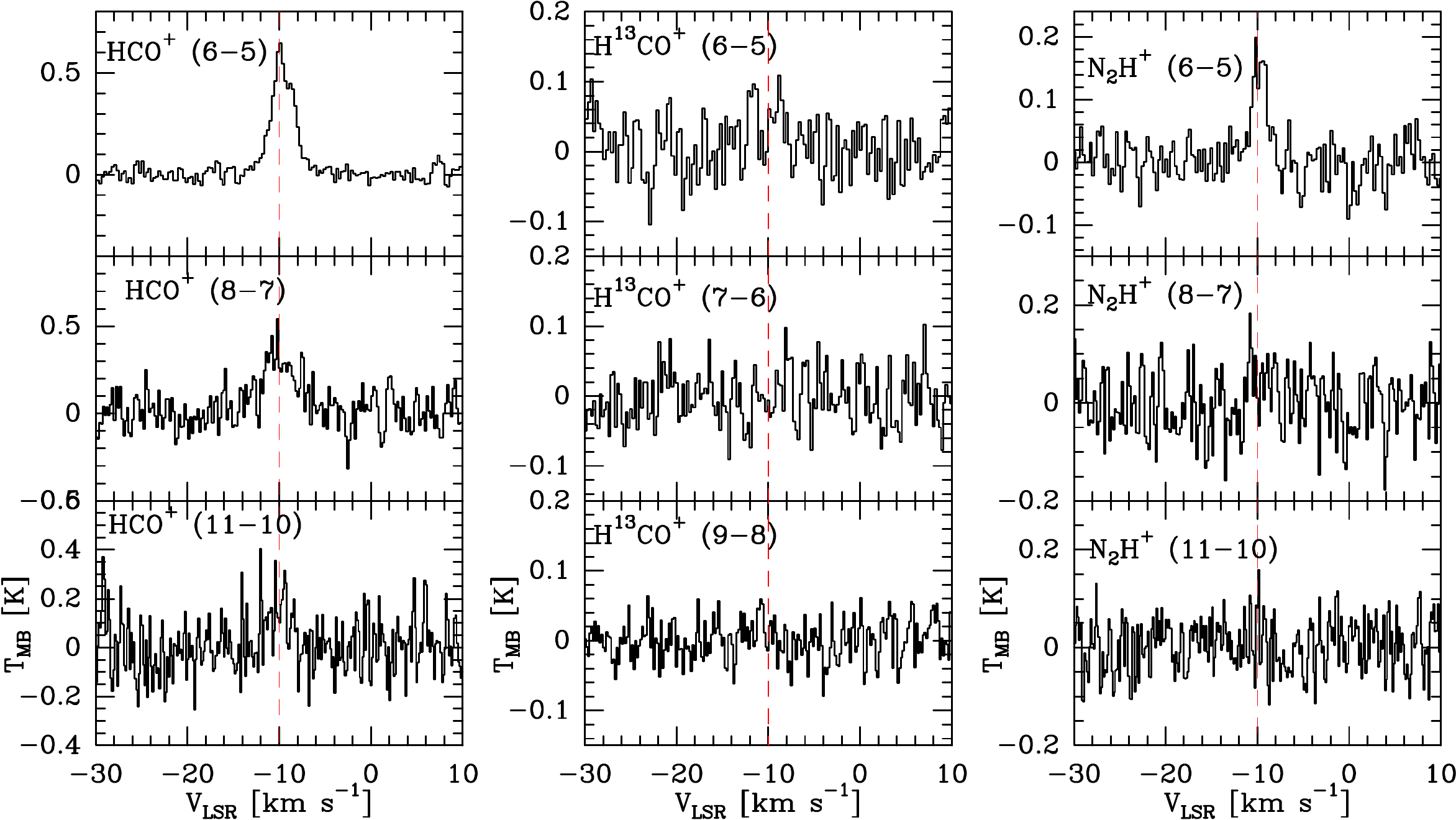}
\caption{Spectra observed toward NGC7129--FIRS2. Dashed red lines indicate a $v_{LSR}$ = $-$9.7~km~s$^{-1}$. The observed transition is indicated on each plot.}
\label{fga7}
\end{figure*}

Figures~\ref{fga1} to~\ref{fga7} display the respective spectra of the HCO$^{+}$, H$^{13}$CO$^{+}$ and N$_{2}$H$^{+}$ transitions observed with Herschel toward a portion of our source sample (see Section~\ref{sec:results}). Note that for display purposes, the spectra have been smoothed at a spectral resolution of 4.4~MHz.

%
%
\section{Transitions of HCO$^{+}$, H$^{13}$CO$^{+}$ and N$_{2}$H$^{+}$ observed with Herschel toward our sample of intermediate and low-mass protostars}
\label{sec:app2}

Tables~\ref{tabB1} to \ref{tabB9} summarize the line parameters for all the HCO$^{+}$, H$^{13}$CO$^{+}$ and N$_{2}$H$^{+}$ transitions observed with Herschel toward our source sample (see Section~\ref{sec:results}).

\begin{table*}
\caption{Line Parameters for the HCO$^{+}$, H$^{13}$CO$^{+}$ and N$_{2}$H$^{+}$ transitions observed toward Serpens--FIRS~1.}\label{tabB1}
\centering
\begin{tabular}{crrrrrrr} 
\hline\hline             
Frequency (MHz)	&	E$\rm_{up}$ (K)& S$\mu$$^{2}$ (D$^{2}$) 	& $v$ (km~s$^{-1}$)& $\Delta$v$_{1/2}$ (km~s$^{-1}$)&T$\rm_{MB}$ (K)&$\int$T$\rm_{MB}$dV (K~km~s$^{-1}$)& Comments\\
(1) & (2) & (3) & (4)& (5) &(6) & (7) & (8)\\
\hline
  \multicolumn{8}{c}{HCO$^{+}$}\\ 
\hline
535061.581	&	89.9	&	90.7	& 8.2(0.1)&2.7(0.2) &1.3(0.1) &3.8(0.6)&D 	\\
713341.228	&	154.1	&	120.9	 & 8.1(0.2)&3.1(0.4) &0.9(0.1)&3.2(0.3)&D	\\
980636.494	&	282.4	&	166.2	& 7.6(0.2)& 3.2(0.5)&0.6(0.4)&1.9(0.3)&	D\\
1069693.891	&	333.8	&	181.3	 & -- &-- &--&--&NO	\\
\hline
  \multicolumn{8}{c}{H$^{13}$CO$^{+}$} \\
\hline
520459.884	&	87.4	&	90.7	&  8.3(0.2)&1.7(0.5) &0.2(0.1)&0.41(0.09)&D	\\
607174.646	&	116.6	&	105.8	& 8.3(0.2)&1.6(0.7) &0.2(0.1)&0.29(0.09)&D	\\
780562.812	&	187.3	&	136.1	& 8.2(0.3) & 0.8(0.5)& 0.09(0.09) &0.08(0.06)& TD	\\

\hline
\multicolumn{8}{c}{N$_{2}$H$^{+}$} \\
\hline
558966.503	&	93.9	&	669.5 & 8.42(0.06)&1.7(0.1) &0.7(0.1)&1.21(0.09)&	D\\
745209.868	&	161.0	&	891.6 & 8.2(0.2)& 2.1(0.4)&0.3(0.2)&0.7(0.1)&D	\\
1024443.025	&	295.1	&	1223.1&   8.3(0.3)& 0.7(0.5)&0.1(0.1)& 0.10(0.08)& TD	\\
\hline
\end{tabular}
\tablefoot{(1)--(3) Spectroscopic line parameters (see references listed in Table~\ref{tab2}) of the observed transitions.
(4)--(7) Observed line parameters: velocity, linewidth, peak temperature and integrated intensities. The 3$\sigma$ uncertainties that result from gaussian fits performed with the CLASS software
are given in brackets.
(8) D: Detected above the 3$\sigma$ level and $\int$T$\rm_{MB}$dV $\ge$ 5$\sigma$, TD: Tentative Detection (line that emits at the 3$\sigma$), NO: Not observed.} 
\end{table*}


\begin{table*}
\caption{Line Parameters for the HCO$^{+}$, H$^{13}$CO$^{+}$ and N$_{2}$H$^{+}$ transitions observed toward IC1396N.}\label{tabB2}
\centering
\begin{tabular}{crrrrrrr} 
\hline\hline             
Frequency (MHz)	&	E$\rm_{up}$ (K)& S$\mu$$^{2}$ (D$^{2}$)	& $v$ (km~s$^{-1}$)& $\Delta$v$_{1/2}$ (km~s$^{-1}$)&T$\rm_{MB}$ (K)&$\int$T$\rm_{MB}$dV (K~km~s$^{-1}$)& Comments\\
(1) & (2) & (3) & (4)& (5) &(6) & (7) & (8)\\
\hline
  \multicolumn{8}{c}{HCO$^{+}$}\\ 
\hline
535061.581	&	89.9	&	90.7	& 0.27(0.03)&2.8(0.1) &2.3(0.1)&6.7(0.6)&D\\
713341.228	&	154.1	&	120.9	& 0.7(0.2)&2.8(0.6) &1.1(0.2) &3.4(0.5)&D	\\
980636.494	&	282.4	&	166.2	& 0.7(0.2)&4.0(0.2) &0.5(0.4)&2.24(0.05)&	D\\
1069693.891	&	333.8	&	181.3	& 0.7(0.2) &3.8(0.2) &0.3(0.2)&1.31(0.03)&	D\\
\hline
  \multicolumn{8}{c}{H$^{13}$CO$^{+}$} \\
\hline
520459.884	&	87.4	&	90.7	& 0.4(0.6)&2.02(1.55) &0.09(0.10)&0.21(0.10)&W	\\
607174.646	&	116.6	&	105.8	& 0.5(0.9) & 2.4(1.7) &0.08(0.08)&0.2(0.1)&TD	\\
780562.812	&	187.3	&	136.1 	& -- &-- &$\le$0.1&$\le$0.4&ND	\\
\hline
\multicolumn{8}{c}{N$_{2}$H$^{+}$} \\
\hline
558966.503	&	93.9	&	669.5 &  0.2(0.1)& 2.4(0.4)&0.4(0.1)&0.9(0.1)&D	\\
745209.868	&	161.0	&	891.6 &    0.01(0.49) & 3.4(1.1) &0.2(0.2)&0.8(0.2)& W	\\
1024443.025	&	295.1	&	1223.1&  -- &-- &$\le$0.2&$\le$0.6&ND	\\
\hline
\end{tabular}
\tablefoot{(1)--(3) Spectroscopic line parameters (see references listed in Table~\ref{tab2}) of the observed transitions.
(4)--(7) Observed line parameters: velocity, linewidth, peak temperature and integrated intensities. The 3$\sigma$ uncertainties that result from gaussian fits performed with the CLASS software
are given in brackets.
(8) D: Detected above the 3$\sigma$ level and $\int$T$\rm_{MB}$dV $\ge$ 5$\sigma$,  W: Weak line ($\int$T$\rm_{MB}$dV $\ge$ 5$\sigma$), ND: Not Detected lines (T$\rm_{MB}$ $<$ 3$\sigma$ and $\int$T$\rm_{MB}$dV $<$ 5$\sigma$), TD: Tentative Detection (line that emits at the 3$\sigma$).} 
\end{table*}

\begin{table*}
\caption{Line Parameters for the HCO$^{+}$, H$^{13}$CO$^{+}$ and N$_{2}$H$^{+}$ transitions observed toward VLA1623.}\label{tabB3}
\centering
\begin{tabular}{crrrrrrr} 
\hline\hline             
Frequency (MHz)	&	E$\rm_{up}$ (K)& S$\mu$$^{2}$ (D$^{2}$) & $v$ (km~s$^{-1}$)& $\Delta$v$_{1/2}$ (km~s$^{-1}$)&T$\rm_{MB}$ (K)&$\int$T$\rm_{MB}$dV (K~km~s$^{-1}$)& Comments\\
(1) & (2) & (3) & (4)& (5) &(6) & (7) & (8))\\
\hline
  \multicolumn{8}{c}{HCO$^{+}$}\\ 
\hline
535061.581	&	89.9	&	90.7	&  3.82(0.04)& 1.2(0.1)& 1.7(0.1)&2.2(0.4)&D 	\\
713341.228	&	154.1	&	120.9	&  3.6(0.1)&0.9(0.3)& 0.8(0.4) &0.7(0.2)&D\\
980636.494	&	282.4	&	166.2	&  -- & -- &$\le$0.5 &$\le$0.5 &ND\\
1069693.891	&	333.8	&	181.3	&  --& --&$\le$0.3&$\le$0.3&ND\\
\hline
  \multicolumn{8}{c}{H$^{13}$CO$^{+}$} \\
\hline
520459.884	&	87.4	&	90.7	&  3.7(0.2)&0.5(0.5) &0.2(0.1)&0.10(0.06) & D\\
607174.646	&	116.6	&	105.8	& --& --&$\le$0.2&$\le$0.2& ND\\
780562.812	&	187.3	&	136.1	& --& --&$\le$0.3&$\le$0.3& ND\\
\hline
\multicolumn{8}{c}{N$_{2}$H$^{+}$} \\
\hline
558966.503	&	93.9	&	669.5 &  3.7(0.1) & 0.7(0.2)&0.2(0.1)&0.16(0.06)&D\\
745209.868	&	161.0	&	891.6 &  --& --&$\le$0.3& $\le$0.3& ND\\
1024443.025	&	295.1	&	1223.1& --& --&$\le$0.2&$\le$0.2& ND\\
\hline
\end{tabular}
\tablefoot{(1)--(3) Spectroscopic line parameters (see references listed in Table~\ref{tab2}) of the observed transitions.
(4)--(7) Observed line parameters: velocity, linewidth, peak temperature and integrated intensities. The 3$\sigma$ uncertainties that result from gaussian fits performed with the CLASS software
are given in brackets.
(8) D: Detected above the 3$\sigma$ level and $\int$T$\rm_{MB}$dV $\ge$ 5$\sigma$, ND: Not Detected lines (T$\rm_{MB}$ $<$ 3$\sigma$ and $\int$T$\rm_{MB}$dV $<$ 5$\sigma$).} 
\end{table*}


\begin{table*}
\caption{Line Parameters for the HCO$^{+}$, H$^{13}$CO$^{+}$ and N$_{2}$H$^{+}$ transitions observed toward L1527.}\label{tabB4}
\centering
\begin{tabular}{crrrrrrr} 
\hline\hline             
Frequency (MHz)	&	E$\rm_{up}$ (K)& S$\mu$$^{2}$ (D$^{2}$)	& $v$ (km~s$^{-1}$)& $\Delta$v$_{1/2}$ (km~s$^{-1}$)&T$\rm_{MB}$ (K)&$\int$T$\rm_{MB}$dV (K~km~s$^{-1}$)& Comments\\
(1) & (2) & (3) & (4)& (5) &(6) & (7) & (8)\\
\hline
  \multicolumn{8}{c}{HCO$^{+}$}\\ 
\hline
535061.581	&	89.9	&	90.7	&  5.52(0.07) & 1.9(0.2) &0.6(0.1) &1.26(0.09) &D 	\\
713341.228	&	154.1	&	120.9	 &-- & --&$\le$0.4&$\le$0.8& ND	\\
980636.494	&	282.4	&	166.2	& -- & --&$\le$0.5&$\le$1.0& ND	\\
1069693.891	&	333.8	&	181.3	&-- & --&$\le$0.3&$\le$0.6& ND	\\
\hline
  \multicolumn{8}{c}{H$^{13}$CO$^{+}$} \\
\hline
520459.884	&	87.4	&	90.7	&  -- &-- &$\le$0.2&$\le$0.3& ND	\\
607174.646	&	116.6	&	105.8	& -- &-- &$\le$0.2&$\le$0.4& ND	\\
780562.812	&	187.3	&	136.1	& -- &-- &$\le$0.2&$\le$0.3& ND	\\
\hline
\multicolumn{8}{c}{N$_{2}$H$^{+}$} \\
\hline
558966.503	&	93.9	&	669.5 & --&-- &$\le$0.2&$\le$0.3&ND	\\
745209.868	&	161.0	&	891.6 &-- & -- &$\le$0.3&$\le$0.5& ND	\\
1024443.025	&	295.1	&	1223.1&-- & -- &$\le$0.2&$\le$0.4& ND	\\
\hline
\end{tabular}
\tablefoot{(1)--(3) Spectroscopic line parameters (see references listed in Table~\ref{tab2}) of the observed transitions.
(4)--(7) Observed line parameters: velocity, linewidth, peak temperature and integrated intensities. The 3$\sigma$ uncertainties that result from gaussian fits performed with the CLASS software
are given in brackets.
(8) D: Detected above the 3$\sigma$ level and $\int$T$\rm_{MB}$dV $\ge$ 5$\sigma$, ND: Not Detected lines (T$\rm_{MB}$ $<$ 3$\sigma$ and $\int$T$\rm_{MB}$dV $<$ 5$\sigma$).} 
\end{table*}


\begin{table*}
\caption{Line Parameters for the HCO$^{+}$, H$^{13}$CO$^{+}$ and N$_{2}$H$^{+}$ transitions observed toward L1157--MM.}\label{tabB5}
\centering
\begin{tabular}{crrrrrrr} 
\hline\hline             
Frequency (MHz)	&	E$\rm_{up}$ (K)& S$\mu$$^{2}$ (D$^{2}$) 	& $v$ (km~s$^{-1}$)& $\Delta$v$_{1/2}$ (km~s$^{-1}$)&T$\rm_{MB}$ (K)&$\int$T$\rm_{MB}$dV (K~km~s$^{-1}$)& Comments\\
(1) & (2) & (3) & (4)& (5) &(6) & (7) & (8)\\
\hline
  \multicolumn{8}{c}{HCO$^{+}$}\\ 
\hline
535061.581	&	89.9	&	90.7	&2.5(0.1) & 1.5(0.2)&0.4(0.1)&0.63(0.09) &D	\\
713341.228	&	154.1	&	120.9	& --& --&$\le$0.5&$\le$0.8& ND	\\
980636.494	&	282.4	&	166.2	& --& --&$\le$0.5&$\le$0.8& ND	\\
1069693.891	&	333.8	&	181.3	& --& --&$\le$0.3&$\le$0.5& ND	\\
\hline
  \multicolumn{8}{c}{H$^{13}$CO$^{+}$} \\
\hline
520459.884	&	87.4	&	90.7	& --& --&$\le$0.1&$\le$0.2& ND	\\
607174.646	&	116.6	&	105.8	& --& --&$\le$0.2&$\le$0.3& ND	\\
780562.812	&	187.3	&	136.1	& --& --&$\le$0.1&$\le$0.2& ND	\\
\hline
\multicolumn{8}{c}{N$_{2}$H$^{+}$} \\
\hline
558966.503	&	93.9	&	669.5 &  --& --&$\le$0.1&$\le$0.2&ND	\\
745209.868	&	161.0	&	891.6 & --& --&$\le$0.3&$\le$0.4& ND\\
1024443.025	&	295.1	&	1223.1& --& --&$\le$0.2&$\le$0.3& ND	\\
\hline
\end{tabular}
\tablefoot{(1)--(3) Spectroscopic line parameters (see references listed in Table~\ref{tab2}) of the observed transitions.
(4)--(7) Observed line parameters: velocity, linewidth, peak temperature and integrated intensities. The 3$\sigma$ uncertainties that result from gaussian fits performed with the CLASS software
are given in brackets.
(8) D: Detected above the 3$\sigma$ level and $\int$T$\rm_{MB}$dV $\ge$ 5$\sigma$, ND: Not Detected lines (T$\rm_{MB}$ $<$ 3$\sigma$ and $\int$T$\rm_{MB}$dV $<$ 5$\sigma$).} 
\end{table*}


\begin{table*}
\caption{Line Parameters for the HCO$^{+}$, H$^{13}$CO$^{+}$ and N$_{2}$H$^{+}$ transitions observed toward NGC1333--IRAS2.}\label{tabB6}
\centering
\begin{tabular}{crrrrrrr} 
\hline\hline             
Frequency (MHz)	&	E$\rm_{up}$ (K)& S$\mu$$^{2}$ (D$^{2}$) 	& $v$ (km~s$^{-1}$)& $\Delta$v$_{1/2}$ (km~s$^{-1}$)&T$\rm_{MB}$ (K)&$\int$T$\rm_{MB}$dV (K~km~s$^{-1}$)& Comments\\
(1) & (2) & (3) & (4)& (5) &(6) & (7) & (8)\\
\hline
  \multicolumn{8}{c}{HCO$^{+}$}\\ 
\hline
535061.581	&	89.9	&	90.7	& 7.52(0.04)&1.2(0.2)&1.03(0.09)&1.4(0.6)&D	\\
713341.228	&	154.1	&	120.9	& 7.6(0.2)& 1.7(0.5)&0.8(0.5)&1.5(0.3)&D	\\
980636.494	&	282.4	&	166.2	& 7.3(0.3)&1.9(0.8) &0.4(0.3)&0.8(0.3)& D	\\
1069693.891	&	333.8	&	181.3	 & -- &-- &--&--&NO	\\
\hline
  \multicolumn{8}{c}{H$^{13}$CO$^{+}$} \\
\hline
520459.884	&	87.4	&	90.7	& 7.5(0.2) &0.7(0.4) &0.1(0.1)&0.09(0.06)&W	\\
607174.646	&	116.6	&	105.8	& -- &-- &$\le$0.1&$\le$0.1&ND	\\
780562.812	&	187.3	&	136.1	& -- &-- &$\le$0.1&$\le$0.1&ND	\\
\hline
\multicolumn{8}{c}{N$_{2}$H$^{+}$} \\
\hline
558966.503	&	93.9	&	669.5 &  7.6(0.2)&1.3(0.4) &0.3(0.2)&0.36(0.09)&D	\\
745209.868	&	161.0	&	891.6 & -- &-- &$\le$0.3&$\le$0.6&ND\\
1024443.025	&	295.1	&	1223.1&  -- &-- &$\le$0.3&$\le$0.6&ND	\\
\hline
\end{tabular}
\tablefoot{(1)--(3) Spectroscopic line parameters (see references listed in Table~\ref{tab2}) of the observed transitions.
(4)--(7) Observed line parameters: velocity, linewidth, peak temperature and integrated intensities. The 3$\sigma$ uncertainties that result from gaussian fits performed with the CLASS software
are given in brackets.
(8) D: Detected above the 3$\sigma$ level and $\int$T$\rm_{MB}$dV $\ge$ 5$\sigma$, W: Weak line ($\int$T$\rm_{MB}$dV $\ge$ 5$\sigma$), ND: Not Detected lines (T$\rm_{MB}$ $<$ 3$\sigma$ and $\int$T$\rm_{MB}$dV $<$ 5$\sigma$), NO: Not observed.} 
\end{table*}


\begin{table*}
\caption{Line Parameters for the HCO$^{+}$, H$^{13}$CO$^{+}$ and N$_{2}$H$^{+}$ transitions observed toward L1641~S3~MMS~1.}\label{tabB7}
\centering
\begin{tabular}{crrrrrrr} 
\hline\hline             
Frequency (MHz)	&	E$\rm_{up}$ (K)& S$\mu$$^{2}$ (D$^{2}$) 	& $v$ (km~s$^{-1}$)& $\Delta$v$_{1/2}$ (km~s$^{-1}$)&T$\rm_{MB}$ (K)&$\int$T$\rm_{MB}$dV (K~km~s$^{-1}$)& Comments\\
(1) & (2) & (3) & (4)& (5) &(6) & (7) & (8)\\
\hline
  \multicolumn{8}{c}{HCO$^{+}$}\\ 
\hline
535061.581	&	89.9	&	90.7	&  5.09(0.07)&1.4(0.3)&0.6(0.1)&0.9(0.4)& D 	\\
713341.228	&	154.1	&	120.9	&4.7(0.4) & 3.1(0.7) &0.5(0.5)&1.5(0.3)& W	\\
980636.494	&	282.4	&	166.2	&4.3(0.4) & 3.7(0.7)&0.4(0.5)&1.6(0.3)&	W\\
1069693.891	&	333.8	&	181.3	&4.7(0.3) & 4.0(0.8)&0.2(0.3)& 1.0(0.3)&	W\\
\hline
  \multicolumn{8}{c}{H$^{13}$CO$^{+}$} \\
\hline
520459.884	&	87.4	&	90.7	&  -- & --&$\le$0.2&$\le$0.2&	ND\\
607174.646	&	116.6	&	105.8	& -- & --&$\le$0.1&$\le$0.1&	ND\\
780562.812	&	187.3	&	136.1	& -- & --&$\le$0.1&$\le$0.1&	ND\\
\hline
\multicolumn{8}{c}{N$_{2}$H$^{+}$} \\
\hline
558966.503	&	93.9	&	669.5 &  5.4(0.2) &0.9(0.6)&0.2(0.1)&0.18(0.09)& D	\\
745209.868	&	161.0	&	891.6 & -- & --&$\le$0.3&$\le$0.3&	ND\\
1024443.025	&	295.1	&	1223.1& -- & --&$\le$0.2&$\le$0.2&	ND\\
\hline
\end{tabular}
\tablefoot{(1)--(3) Spectroscopic line parameters (see references listed in Table~\ref{tab2}) of the observed transitions.
(4)--(7) Observed line parameters: velocity, linewidth, peak temperature and integrated intensities. The 3$\sigma$ uncertainties that result from gaussian fits performed with the CLASS software
are given in brackets.
(8) D: Detected above the 3$\sigma$ level and $\int$T$\rm_{MB}$dV $\ge$ 5$\sigma$, W: Weak line ($\int$T$\rm_{MB}$dV $\ge$ 5$\sigma$), ND: Not Detected lines (T$\rm_{MB}$ $<$ 3$\sigma$ and $\int$T$\rm_{MB}$dV $<$ 5$\sigma$).}
\end{table*}


\begin{table*}
\caption{Line Parameters for the HCO$^{+}$, H$^{13}$CO$^{+}$ and N$_{2}$H$^{+}$ transitions observed toward Cep~E--mm.}\label{tabB8}
\centering
\begin{tabular}{crrrrrrr} 
\hline\hline             
Frequency (MHz)	&	E$\rm_{up}$ (K)& S$\mu$$^{2}$ (D$^{2}$)& $v$ (km~s$^{-1}$)& $\Delta$v$_{1/2}$ (km~s$^{-1}$)&T$\rm_{MB}$ (K)&$\int$T$\rm_{MB}$dV (K~km~s$^{-1}$)& Comments\\
(1) & (2) & (3) & (4)& (5) &(6) & (7) & (8)\\
\hline
  \multicolumn{8}{c}{HCO$^{+}$}\\ 
\hline
535061.581	&	89.9	&	90.7	& $-$11.22(0.09) &2.6(0.3) &0.7(0.2)&1.9(0.2)&D	\\
713341.228	&	154.1	&	120.9	& $-$10.9(0.4)&2.9(0.9) &0.4(0.4)&1.3(0.4)&W	\\
980636.494	&	282.4	&	166.2	& $-$11.5(0.2)& 0.5(0.5)&0.5(0.5)&0.3(0.2)&TD	\\
1069693.891	&	333.8	&	181.3	& --& --&$\le$0.3&$\le$0.9&ND	\\
\hline
  \multicolumn{8}{c}{H$^{13}$CO$^{+}$} \\
\hline
520459.884	&	87.4	&	90.7	& --& --&$\le$0.2&$\le$0.2&ND	\\
607174.646	&	116.6	&	105.8	& --& --&$\le$0.2&$\le$0.2&ND	\\
780562.812	&	187.3	&	136.1	& --& --&$\le$0.2&$\le$0.2&ND	\\
\hline
\multicolumn{8}{c}{N$_{2}$H$^{+}$} \\
\hline
558966.503	&	93.9	&	669.5 &  $-$10.9(0.2)& 0.6(0.4)&0.2(0.1)&0.11(0.06)&W	\\
745209.868	&	161.0	&	891.6 &  -- &-- &$\le$0.3&$\le$0.3&ND	\\
1024443.025	&	295.1	&	1223.1&-- &-- &$\le$0.2&$\le$0.2&ND	\\
\hline
\end{tabular}
\tablefoot{(1)--(3) Spectroscopic line parameters (see references listed in Table~\ref{tab2}) of the observed transitions.
(4)--(7) Observed line parameters: velocity, linewidth, peak temperature and integrated intensities. The 3$\sigma$ uncertainties that result from gaussian fits performed with the CLASS software
are given in brackets.
(8) D: Detected above the 3$\sigma$ level and $\int$T$\rm_{MB}$dV $\ge$ 5$\sigma$, W: Weak line ($\int$T$\rm_{MB}$dV $\ge$ 5$\sigma$), ND: Not Detected lines (T$\rm_{MB}$ $<$ 3$\sigma$ and $\int$T$\rm_{MB}$dV $<$ 5$\sigma$) and, TD: Tentative Detection (line that emits at the 3$\sigma$).} 
\end{table*}


\begin{table*}
\caption{Line Parameters for the HCO$^{+}$, H$^{13}$CO$^{+}$ and N$_{2}$H$^{+}$ transitions observed toward NGC7129--FIRS2.}\label{tabB9}
\centering
\begin{tabular}{crrrrrrr} 
\hline\hline             
Frequency (MHz)	&	E$\rm_{up}$ (K)& S$\mu$$^{2}$ (D$^{2}$) 	& $v$ (km~s$^{-1}$)& $\Delta$v$_{1/2}$ (km~s$^{-1}$)&T$\rm_{MB}$ (K)&$\int$T$\rm_{MB}$dV (K~km~s$^{-1}$)& Comments\\
(1) & (2) & (3) & (4)& (5) &(6) & (7) & (8)\\
\hline
  \multicolumn{8}{c}{HCO$^{+}$}\\ 
\hline
535061.581	&	89.9	&	90.7	& $-$9.7(0.1)& 2.5(0.5)&0.46(0.09)&1.2(0.2)&D	\\
713341.228	&	154.1	&	120.9	& $-$9.9(0.6)& 4.8(1.6)&0.3(0.4)&1.6(0.4)&W	\\
980636.494	&	282.4	&	166.2	& $-$10.1(0.7)& 2.8(1.5)&0.2(0.4)&0.5(0.3)&W	\\
1069693.891	&	333.8	&	181.3	& --& --&--&--&	NO\\
\hline
  \multicolumn{8}{c}{H$^{13}$CO$^{+}$} \\
\hline
520459.884	&	87.4	&	90.7	& -- & --&$\le$0.2&$\le$0.5&ND	\\
607174.646	&	116.6	&	105.8	&  -- & --&$\le$0.2&$\le$0.5&ND	\\
780562.812	&	187.3	&	136.1	&  -- & --&$\le$0.1&$\le$0.4&ND	\\
\hline
\multicolumn{8}{c}{N$_{2}$H$^{+}$} \\
\hline
558966.503	&	93.9	&	669.5 &$-$9.8(0.2) &1.9(0.6) &0.2(0.1)&0.34(0.09)&D	\\
745209.868	&	161.0	&	891.6 &  -- & --&$\le$0.3&$\le$0.6&ND	\\
1024443.025	&	295.1	&	1223.1& -- & --&$\le$0.2&$\le$0.4&ND	\\
\hline
\end{tabular}
\tablefoot{(1)--(3) Spectroscopic line parameters (see references listed in Table~\ref{tab2}) of the observed transitions.
(4)--(7) Observed line parameters: velocity, linewidth, peak temperature and integrated intensities. The 3$\sigma$ uncertainties that result from gaussian fits performed with the CLASS software
are given in brackets.
(8) D: Detected above the 3$\sigma$ level and $\int$T$\rm_{MB}$dV $\ge$ 5$\sigma$, ND: Not Detected lines (T$\rm_{MB}$ $<$ 3$\sigma$ and $\int$T$\rm_{MB}$dV $<$ 5$\sigma$), W: Weak line ($\int$T$\rm_{MB}$dV $\ge$ 5$\sigma$), NO: Not observed.} 
\end{table*}
%

\end{appendix}

%
%
\end{document}